\documentclass[aps,prd,preprint,showpacs,showkeys]{revtex4}
\usepackage{graphics}
\newcommand{\V}{\mathcal{V}}
\newcommand{\e}{\mathrm{e}}
\newcommand{\ep}{\epsilon}
\newcommand{\vev}[1]{\left\langle ~#1~ \right\rangle}
\newcommand{\tpt}{(t+t')}
\newcommand{\emi}{\e^{-(t+t')}}
\begin{document}
\preprint{KOBE-TH-03-02} \title{Beta Functions in the Integral
Equation Approach \\to the Exact Renormalization Group}
\author{Hidenori Sonoda} \email[E-mail
address:~]{sonoda@phys.sci.kobe-u.ac.jp} \affiliation{Physics
Department, Kobe University, Kobe 657-8501, Japan} \date{\today}
\begin{abstract}
We incorporate running parameters and anomalous dimensions into the
framework of the exact renormalization group.  We modify the exact
renormalization group differential equations for a real scalar field
theory, using the anomalous dimensions of the squared mass and the
scalar field.  Following a previous paper in which an integral
equation approach to the exact renormalization group was introduced,
we reformulate the modified differential equations as integral
equations that define the continuum limit directly in terms of a
running squared mass and self-coupling constant.  Universality of the
continuum limit under an arbitrary change of the momentum cutoff
function is discussed using the modified exact renormalization group
equations.
\end{abstract}
\pacs{11.10.Gh, 05.10.Cc, 11.10.Hi}
\keywords{exact renormalization group, beta functions, anomalous dimensions}
\maketitle

\section{\label{sec:intro}Introduction}

The renormalization group is a key concept in modern quantum field
theory.  However, we have two drastically different approaches to the
renormalization group.  One gives the familiar scale dependence of the
coupling constants for renormalizable field theories.  The other is
the exact renormalization group (ERG) of Wilson \cite{wk74} in which
all possible interactions are considered within a fixed cutoff scheme.
The two approaches differ in the number of parameters that we must
retain.  In the former case, only renormalizable field theories are
considered, and they are described by a finite number of relevant (and
marginal) parameters.  In the latter case, however, we must keep track
of an infinite number of terms in the action as we change the energy
scale of the theory.  It is the purpose of the present paper to unite
the two approaches.

It is clear what we must do.  We must first restrict the application
of the ERG to the continuum limit, i.e., already renormalized
theories.  The continuum limit constitutes a finite dimensional
subspace of the theory space, and it is closed under the application
of the ERG.  The ordinary renormalized parameters can be interpreted
as coordinates of the finite dimensional subspace.  We expect that the
action of the ERG reduces to the usual scale dependence of the
renormalized parameters.

The continuum limit is usually obtained by taking a certain limit
(such as taking the momentum cutoff to infinity) of a bare theory.
The necessity of a bare theory is obviously a nuisance when our main
interest is in seeing how the ERG acts on the continuum limit rather
than seeing how the limit is approached.  It is much more preferable
that we have direct access to the continuum limit.

In a previous paper \cite{s02} we have introduced an integral equation
approach to the ERG, in terms of which we have constructed the
continuum limit directly without starting from any bare theory.  In
particular it has been shown that the continuum limit of the
perturbative $\phi^4$ theory has \textbf{four} parameters: a squared
mass $m^2$ in the free propagator and three constants of integration
left undetermined when we convert the ERG differential equations into
ERG integral equations.

We know that the continuum limit of the $\phi^4$ theory has only two
parameters, not four.  Our first task in the paper is to reduce the
four parameters to two by removing two redundant degrees of freedom.
It might be possible to impose two conditions on the interaction
vertices so that the conditions are preserved under the ERG flow.
This would relate two parameters to the other two, effectively
reducing the dimensionality of the continuum limit by two.  This is
not, however, the approach we take in this paper.  We will impose two
conditions on the interaction vertices that are easy to enforce.
Unfortunately we will find that the two conditions are not preserved
under the ERG transformation.  To preserve the conditions, we are
forced to modify the ERG transformation itself.  This modification
essentially consists of moving part of the squared mass from the free
part of the action to the interaction part and of changing the
normalization of the scalar field, so that the physical content of the
interaction vertices is kept intact.

The continuum limit now has only two physical parameters: a squared
mass and a self-coupling constant.  Our modified ERG shares two nice
properties with the minimal subtraction scheme in dimensional
regularization \cite{th73}:
\begin{enumerate}
\item \textbf{mass independence}: the squared mass $m^2$ gets renormalized
multiplicatively, and the massless theory corresponds to $m^2 = 0$.
\item \textbf{beta functions determine the subtractions}: the
subtractions necessary to get ultraviolet finiteness are completely
determined by the beta function and anomalous dimensions.
\end{enumerate}

How to derive running renormalized parameters from the ERG has been
considered before.  In Ref.~\onlinecite{hl88} Hughes and Liu have
attempted to introduce a beta function and anomalous dimensions for
the four dimensional $\phi^4$ theory.  Their discussion was incomplete
mainly because of the lack of direct access to the continuum limit,
and partially because of the insufficient care paid to the mass and
wave function counterterms.

The present paper is organized as follows.  In sect.~\ref{sec:review}
we review the ERG and the results of Ref.~\onlinecite{s02} for the
convenience of the reader.  In sect.~\ref{sec:counter} we discuss
carefully the counterterms for the squared mass and wave function, to
prepare for the modification of the ERG in the next section.  In
sect.~\ref{sec:modify} we will modify the ERG differential equation so
as to preserve the two conditions that we impose on the interaction
vertices.  In sect.~\ref{sec:integral} we reformulate the modified ERG
equation as an integral equation, and as a result we identify the two
parameters of the continuum limit.  We will also show that the
subtractions necessary for ultraviolet finiteness are determined by
the beta function and anomalous dimensions.  In sect.~\ref{sec:linear}
we generalize the counterterms introduced in sect.~\ref{sec:counter}
to prepare for the discussion of universality in sect.~\ref{sec:univ}.
Universality has been discussed in Ref.~\onlinecite{s02} in the
context of the original ERG equation.  More details are provided to
supplement the discussion given for the original ERG equation in
Ref.~\onlinecite{s02}.  Finally, we give concluding remarks in
sect.~\ref{sec:conclusion}.

We have aimed at a clear presentation, and the main text has become
inevitably long even without showing some useful details, which we
have collected in the appendices.  Appendices \ref{appendix:beta} and
\ref{appendix:lowest} supplement the discussion in the main text by
showing concrete calculations.  In Appendix \ref{appendix:improvement}
we complete what Hughes and Liu attempted in Ref.~\onlinecite{hl88}.
In Appendix \ref{appendix:SSB} we discuss the case of a negative
squared mass corresponding to spontaneous breaking of the
$\mathbf{Z_2}$ symmetry.  The appendices can be skipped in the initial
reading of the paper.

\section{\label{sec:review}Review: ERG and the integral equation approach}

In this section we summarize the exact renormalization group (ERG) of
Wilson as formulated by Polchinski \cite{pol84} in a form convenient
for perturbation theory.  We also sketch the essential results of the
previous paper \cite{s02} on which the present paper depends heavily.
To contain the section to a reasonable length, the sketch is crude,
and we must refer the reader to Ref.~\onlinecite{s02} for more
details.  For a general review on the subject of ERG, we refer the
reader to Ref.~\onlinecite{bb01}, and for the continuum limit (a.k.a.,
``perfect actions'') to Ref.~\onlinecite{has98}.

We consider the $\mathbf{Z_2}$ invariant $\phi^4$ theory in euclidean
four dimensions.  The theory is defined perturbatively by the full
action
\begin{equation}
S[\phi] =  \int_p \frac{1}{2} \phi (p) \phi (-p) \, \frac{p^2 +
m^2}{K(p)} + S_{\text{int}} [\phi]
\end{equation}
where $\phi (p)$ is the Fourier transform of the scalar field $\phi$,
and the momentum cutoff function $K(p)$ is a smooth non-negative
function of $p^2$ with the property\footnote{$K(p)$ does not have to
vanish exactly for $p^2 > 2^2$.  It only needs to decrease reasonably
fast as $p^2 \to \infty$.  We take $K(p)$ independent of the squared
mass $m^2$.}
\begin{equation}
K(p) = \left\{ \begin{array} {r@{\quad\text{if}\quad}l} 1 & p^2 <
1\\ 0 & p^2 > 2^2 \end{array} \right.
\end{equation}
The interaction action is expanded in powers of the field variables as
\begin{equation}
S_{\text{int}} [\phi] = - \sum_{n=1}^\infty \frac{1}{(2n)!}
\int_{p_1,\cdots, p_{2n-1}} \phi (p_1) \cdots \phi (p_{2n-1})\,
\V_{2n} (p_1, \cdots, p_{2n}) \label{expansion}
\end{equation}
where the total momentum is conserved:
\begin{equation}
p_1 + \cdots + p_{2n} = 0,
\end{equation}
and the integral is taken over $2n-1$ independent momenta.  We call
the coefficients $\V_{2n}~(n=1,\cdots,\infty)$ as interaction
vertices.

The free part of the action determines the propagator as
\begin{equation}
\frac{K(p)}{p^2 + m^2}
\end{equation}
so that the $2n$-point Green functions are computed as
\begin{eqnarray}
&&\vev{ \phi (p_1) \cdots \phi (p_{2n-1}) \phi }_{m^2; \V} \equiv \int
[d\phi] \phi (p_1) \cdots \phi (p_{2n-1}) \phi \: \e^{- S[\phi]}
\nonumber\\&&\quad = \exp \left[ \frac{1}{2} \int_p
\frac{K(p)}{p^2+m^2} \frac{\delta}{\delta \phi (p)}
\frac{\delta}{\delta \phi (-p)} \right] \phi (p_1) \cdots \phi
(p_{2n-1}) \phi \: \e^{- S_{\text{int}} [\phi]} \Bigg|_{\phi = 0}
\end{eqnarray}
Since the propagator vanishes for high momentum, all loop integrals and
therefore the Green functions are finite in perturbation theory.

Given a squared mass $m^2$ and an interaction action
$S_{\text{int}}[\phi]$ (or equivalently the vertices $\{\V_{2n}
(p_1,\cdots,p_{2n})\}$), we can generate an ERG trajectory
parametrized by a logarithmic scale variable $t$.  At each $t$, we
have a theory with a squared mass $m^2 \e^{2t}$ and an interaction
action $S_{\text{int}}[t; \phi]$ (vertices
$\{\V_{2n}(t;p_1,\cdots,p_{2n})\}$) so that the Green functions are
related to the original theory (at $t=0$) as \footnote{Strictly
speaking, this is valid only if $p_i^2 < 1$ for all $i$.  But this is
an inessential technicality.}
\begin{equation}
\e^{(4n - y_{2n})t} \vev{ \phi (p_1 \e^t) \cdots \phi (p_{2n-1}
\e^t) \phi }_{m^2 \e^{2t}; \V(t)} = \vev{ \phi (p_1) \cdots
\phi (p_{2n-1}) \phi}_{m^2; \V}
\end{equation}
where
\begin{equation}
y_{2n} \equiv 4 - 2n
\end{equation}
Physically the theory at $t$ is related to the theory at $t=0$ via a
scale transformation by a factor of $\e^t$.

The $t$-dependence of the interaction action is given by the following
ERG equation\cite{wk74,pol84}:
\begin{equation}
- \frac{\partial}{\partial t} S_{\text{int}} [t; \phi] = \frac{1}{2}
  \int_p \frac{\Delta (p)}{p^2 + m^2 \e^{2t}} \left( \frac{\delta
  S_{\text{int}}[t;\phi]}{\delta \phi (p)} \frac{\delta S_{\text{int}}
  [t;\phi]}{\delta \phi (-p)} - \frac{\delta^2 S_{\text{int}}
  [t;\phi]}{\delta \phi (p) \delta \phi (-p)} \right)
\end{equation}
where
\begin{equation}
\Delta (p) \equiv - 2 p^2 \frac{d}{d p^2} K(p) \label{Delta}
\end{equation}
is non-vanishing only for $1 < p^2 < 2^2$.  Substituting the expansion
(\ref{expansion}) into the above, we obtain the ERG differential
equations for the vertices:
\begin{eqnarray}
&&\frac{\partial}{\partial t} \left( \e^{- y_{2n} t} \V_{2n} (t; p_1
\e^t, \cdots, p_{2n} \e^t) \right) \nonumber\\
&=& \sum_{k=0}^{\left[\frac{n-1}{2}\right]} \sum_{\text{partitions}
\atop I+J = \{2n\}} \e^{- y_{2(k+1)}t} \V_{2(k+1)} (t; p_I \e^t)\,
\frac{\Delta (p_I \e^t)}{p_I^2 + m^2}\, \e^{- y_{2(n-k)}t} \V_{2(n-k)}
(t; p_J \e^t)\nonumber\\ && + \frac{1}{2} \int_q \frac{\Delta (q \e^t)}{q^2
+ m^2} \, \e^{- y_{2(n+1)}t} \V_{2(n+1)} (t; q \e^t, - q \e^t, p_1 \e^t,
\cdots, p_{2n} \e^t)
\end{eqnarray}
where the Gauss symbol $[x]$ denotes the largest integer not bigger
than $x$, and the second sum in the double summation is over all
possible ways of partitioning $2n$ external momenta
$p_1,\cdots,p_{2n}$ into two groups of $2k+1$ and $2(n-k) -1$
elements.  The notation $p_I$ is a shorthand for either a list of
$2k+1$ momenta or their sum.\footnote{If we use a fuller notation,
$p_I$ should be either a set $\{p_{I_1},\cdots,p_{I_{2k+1}}\}$ or
their sum, where $I_1,\cdots,I_{2k+1}$ are $2k+1$ elements from $1$ to
$2n$.  The notation $\V_{2(k+1)} (p_I)$ should be interpreted as
$\V_{2(k+1)} \left(
p_{I_1},\cdots,p_{I_{2k+1}},-(p_{I_1}+\cdots+p_{I_{2k+1}})\right)$.
We often omit the last argument, since it can be implied by momentum
conservation.  Hence, we often write $\V_2 (p)$ instead of $\V_2
(p,-p)$.}  The same goes for $p_J$.

We can expand the interaction vertices in powers of $m^2$:
\begin{equation}
\V_{2n} (t; p_1,\cdots,p_{2n}) = A_{2n} (t; p_1,\cdots,p_{2n}) + m^2 B_{2n}
(t; p_1,\cdots,p_{2n}) + \text{O} (m^4)
\end{equation}
In particular we introduce the four coefficients at zero momenta:
\begin{eqnarray}
A_2 (t) &\equiv& A_2 (t; 0,0)\\
B_2 (t) &\equiv& B_2 (t; 0,0)\\
C_2 (t) &\equiv& \frac{\partial}{\partial p^2} A_2 (t; p,-p)\Big|_{p^2=0}\\
A_4 (t) &\equiv& A_4 (t; 0,0,0,0)
\end{eqnarray}

In Ref.~\onlinecite{s02} we have reformulated the above ERG
differential equations as integral equations that define the continuum
limit directly.  In particular we have shown that the continuum limit
is parametrized by the four parameters $m^2$, $B_2 (0)$, $C_2 (0)$,
and $A_4 (0)$.  Given these four parameters, the integral equations
determine the entire ERG trajectory from $t=-\infty$ to $t=0$, and
therefore we can regard the four parameters as the coordinates of the
end point $t=0$ of the ERG trajectory.  For example, the four-point
vertex is determined by the ERG integral equation as
follows\footnote{We have replaced $t$ by $-t$ so that $t$ becomes
positive along the ERG trajectory.  The left-hand side should be $\V_4
(-t; p_1,\cdots,p_4)$ for arbitrary $t > 0$, but we have only written
down the equation for $t=0$ for simplicity.}:
\begin{eqnarray}
&&\V_4 (0; p_1,\cdots, p_4) \nonumber\\ &=& \int_0^\infty dt \, \Bigg[
\, \sum_{i=1}^4 \e^{2 t} \V_2 (-t; p_i \e^{-t}) \frac{\Delta (p_i
\e^{-t})}{p_i^2 + m^2} \, \V_4 (-t; p_1,\cdots,p_4)\nonumber\\ &&
\quad + \frac{1}{2} \int_q \Bigg\lbrace \frac{\Delta (q \e^{-t})}{q^2
+ m^2} \, \e^{- 2t} \V_6 (-t; q \e^{-t}, - q \e^{-t}, p_1
\e^{-t},\cdots,p_4 \e^{-t}) \nonumber\\
&&\qquad\qquad - \frac{\Delta (q \e^{-t})}{q^2} \e^{-2 t}
A_6 (-t; q \e^{-t}, -q \e^{-t},0,0,0,0) \Bigg\rbrace \,\Bigg]
\: + A_4 (0)
\end{eqnarray}
Similarly, the integral equation for $\V_2$ depends on $B_2 (0)$, $C_2
(0)$.  But the integral equations for $\V_{2n \ge 6}$ do not depend on
any of the parameters explicitly.  The integral equations are far from
mutually independent.  $\V_{2n \ge 6}$ depend implicitly on all the
parameters, since $\V_{2,4}$ enter into the integral equations
determining $\V_{2n \ge 6}$.  

The biggest advantage of the formulation in terms of the integral
equations is that they define the continuum limit directly.  Solved
recursively, the integral equations naturally reproduce perturbation
theory.  See sect.~IV of Ref.~\onlinecite{s02} for more details.

Our task is to reduce the number of free parameters from four to two.
The next section prepares ourselves for that task.

\section{\label{sec:counter}Counterterms for the squared mass and wave
function}

We will eventually modify the ERG equations by adding ``counterterms''
to the vertices.  It is straightforward to introduce counterterms to a
bare action in a regularization scheme such as the dimensional
regularization.  In our case, however, introduction of counterterms
needs some care.  Naively we would change the squared mass as
\begin{equation}
m^2 \longrightarrow m^2 + \Delta m^2
\end{equation}
and compensate this change by another change
\begin{equation}
\V_2 (p) \longrightarrow \V_2 (p) + \Delta m^2
\end{equation}
so that the total action remains invariant.  This is essentially
correct, but not quite so in this case.  Strictly speaking, the mass
term is not $m^2$, but it is divided by the cutoff function $K(p)$ as
$m^2/K(p)$.  The above two changes do not keep the total action
invariant.  We must examine the notion of counterterms more carefully.

The full action is given by
\begin{eqnarray}
S[\phi] &=& \int_p \frac{1}{2} \left( \frac{p^2 + m^2}{K(p)} - \V_2
(p) \right) \phi (p) \phi (-p) \nonumber\\ && - \sum_{n=2}^\infty
\frac{1}{(2n)!} \int_{p_1,\cdots,p_{2n-1}} \V_{2n}
(p_1,\cdots,p_{2n})\, \phi(p_1) \cdots \phi (p_{2n}) \label{full}
\end{eqnarray}
The part quadratic in the field consists of the free part that
determines the propagator and of the interaction part $\V_2$.  This
splitting is not uniquely determined and susceptible to a convention.
In the following we will describe a correct way of adding counterterms
for the squared mass and wave function without changing the physical
content of the theory.  More general counterterms, necessary for the
discussion of universality, will be introduced later in
sect.~\ref{sec:linear}.

To introduce counterterms, we start with an infinitesimal change of
field variables.  In the full action we make the following replacement
of the field:
\begin{equation}
\phi (p) \longrightarrow \phi (p) \left( 1 + \frac{1}{2} s(p) \right)
\label{change}
\end{equation}
where $s(p)$ is an infinitesimal function of $p^2$.  This replacement
changes the action to
\begin{eqnarray}
S'[\phi] &=& \int_p \frac{1}{2} \phi (p) \phi (-p) \left(
\frac{p^2+m^2}{K(p)} - \V_2 (p) \right) \left( 1 + s (p)
\right)\nonumber\\
&& - \sum_{n=2}^\infty \frac{1}{(2n)!} \int_{p_1,\cdots,p_{2n-1}}
\V_{2n} (p_1,\cdots,p_{2n})\, \phi (p_1) \cdots \phi (p_{2n}) \left( 1
+ \frac{1}{2} \sum_{i=1}^{2n} s(p_i) \right) 
\end{eqnarray}
The coefficient of the quadratic part can be rewritten as follows:
\begin{eqnarray}
\left( \frac{p^2 + m^2}{K(p)} - \V_2
(p) \right) (1 + s(p)) &=& \frac{p^2 + m^2 (1+\ep)}{K(p)}
- \ep m^2 - \eta (p^2 + m^2) - \V_2 (p) (1 + s(p)) \nonumber\\ &&\quad
+~ \ep m^2 \left(1 - \frac{1}{K(p)}\right) + \left(\frac{s(p)}{K(p)} +
\eta\right) (p^2+m^2)
\end{eqnarray}
where $\ep$, $\eta$ are arbitrary infinitesimal constants.  The second
line of the right-hand side vanishes if we choose
\begin{equation}
s(p) = (1 - K(p)) \frac{\ep m^2}{p^2 + m^2} - \eta K(p) = - \eta + (1
- K(p)) \left( \eta + \frac{\ep m^2}{p^2 + m^2} \right) \label{choice}
\end{equation}
Note that $s(p)$ is a smooth function of $p^2$, and 
\begin{equation}
s(p) = - \eta \quad \text{for}\quad p^2 < 1 \label{seta}
\end{equation}
since $K(p) = 1$ for $p^2 < 1$.

With the above choice for $s(p)$, we obtain
\begin{eqnarray}
S' [\phi] &\equiv& \int_p \frac{1}{2} \frac{p^2 + m^2 (1+\ep)}{K(p)}
\phi(p) \phi (-p) \nonumber\\ && - \sum_{n=1}^\infty \frac{1}{(2n)!}
\int_{p_1,\cdots,p_{2n-1}} \V'_{2n} (p_1,\cdots,p_{2n})\, \phi(p_1)
\cdots \phi (p_{2n}) \label{Sprime}
\end{eqnarray}
where
\begin{eqnarray}
\V'_2 (p) &=& \ep m^2 + \eta (p^2 + m^2) + (1 + s(p)) \V_2 (p)
\label{Vtwoprime}\\ \V'_{2n \ge 4} (p_1, \cdots, p_{2n}) &=& \V_{2n} (p_1,
\cdots, p_{2n}) \left( 1 + \frac{1}{2} \sum_{i=1}^{2n} s(p_i) \right)
\label{Vprime}
\end{eqnarray}

We make the following observations:
\begin{enumerate}
\item The new theory has the squared mass $m^2 (1 + \ep)$, and the
propagator is given by
\begin{equation}
\frac{K(p)}{p^2 + m^2 (1+\ep)}
\end{equation}
\item If $s (p) = - \eta$, Eqs.~(\ref{Vtwoprime}, \ref{Vprime}) would
be nothing but the addition of the naive counterterms.
\item For $p^2 > 1$, $s(p)$ is momentum dependent, and it depends also
on the mass shift.
\item The new theory is physically equivalent to the original, since
they are related simply by an infinitesimal linear change of field
variables.  More specifically, the Green functions are related by
\begin{equation}
\vev{\phi (p_1) \cdots \phi (p_{2n-1}) \phi}_{m^2; \V}
= (1 - n \eta) \vev{\phi (p_1) \cdots \phi (p_{2n-1}) \phi}_{m^2
(1+\ep); \V'}
\end{equation}
if $p_i^2 < 1$ for all $i$.
\end{enumerate}

We now have everything we need in order to modify the ERG equation.

\section{\label{sec:modify}Modification of the ERG equations}

The counterterms introduced in the previous section has two arbitrary
infinitesimal constants $\ep$ and $\eta$.  Expanding the two-point
vertices in powers of $m^2$, Eq.~(\ref{Vtwoprime}) gives
\begin{eqnarray}
\frac{d}{dp^2} A'_2 (p)\Big|_{p^2=0} &=& \eta + (1 - \eta)
\frac{d}{dp^2} A_2 (p)\Big|_{p^2=0} \\
B'_2 (p=0) &=& \ep + \eta + (1 - \eta) B_2 (0)
\end{eqnarray}
This suggests that using the counterterms we can modify the vertex
functions so that the two conditions
\begin{equation}
\frac{d}{dp^2} A'_2 (p)\Big|_{p^2=0} = B'_2 (p=0) = 0
\end{equation}
are satisfied.  As we renormalize the vertices, we must keep adding
counterterms so that the above conditions are satisfied along the
entire renormalization group trajectory.  The ERG differential
equation will be modified so as to include the necessary counterterms
to preserve the above conditions.

Let us start with the vertices $\{\V_{2n}\}$ satisfying the conditions
\begin{equation}
\frac{d}{dp^2} A_2 (p)\Big|_{p^2=0} = B_2 (p=0) = 0 \label{conditions}
\end{equation}
We then renormalize them by an infinitesimal logarithmic scale $\Delta
t$ to obtain the new vertices $\{\tilde{\V}_{2n}\}$ given by
\begin{eqnarray}
&&\e^{- y_{2n} \Delta t} \tilde{\V}_{2n} (p_1 \e^{\Delta t},\cdots, p_{2n}
\e^{\Delta t}) - \V_{2n} (p_1,\cdots,p_{2n}) \nonumber\\ &=& \Delta t
\Bigg[ \, \sum_{k=0}^{\left[\frac{n-1}{2}\right]}
\sum_{\text{partitions}\atop I+J=\{2n\}} \V_{2(k+1)} (p_I)
\frac{\Delta (p_I)}{p_I^2 + m^2} \V_{2(n-k)} (p_J)\nonumber\\ && \qquad
+ \frac{1}{2} \int_q \frac{\Delta (q)}{q^2+m^2} \, \V_{2(n+1)}
(q,-q,p_1,\cdots,p_{2n}) \,\Bigg]
\end{eqnarray}
Especially for the two-point vertex, this implies
\begin{eqnarray}
\frac{d}{dp^2} \tilde{A}_2 (p)\Big|_{p^2=0} &=& \Delta t
\frac{\partial}{\partial p^2} \left( \frac{1}{2} \int_q \frac{\Delta
(q)}{q^2} A_4 (q,-q,p,-p) \right)_{p^2=0} \label{tildeCtwo}\\
\tilde{B}_2 (0) &=& \Delta t \frac{1}{2} \int_q \Delta (q) \left(
\frac{1}{q^2} B_4 (q,-q,0,0) - \frac{1}{q^4} A_4 (q,-q,0,0) \right)
\label{tildeBtwo} 
\end{eqnarray}
where we used the conditions (\ref{conditions}).  Hence, the
transformed vertices $\{\tilde{\V}_{2n}\}$ do not satisfy the
vanishing conditions (\ref{conditions}) anymore.

We wish to modify the vertices $\{\tilde{\V}_{2n}\}$ into an
equivalent set of vertices $\{\V''_{2n}\}$ that satisfy the conditions
(\ref{conditions}).  The necessary modification can be done using the
counterterms derived in the previous section.  In terms of two
infinitesimal constants $\ep$, $\eta$, the modified vertices are given
by
\begin{eqnarray}
\V''_2 (p) &=& \ep m^2 + \eta (p^2 + m^2) + (1 + s(p)) \tilde{\V}_2
(p)\\
\V''_{2n \ge 4} (p_1,\cdots,p_{2n}) &=& \left( 1 + \frac{1}{2}
\sum_{i=1}^{2n} s(p_i) \right) \tilde{\V}_{2n} (p_1,\cdots,p_{2n})
\end{eqnarray}
where $s(p)$ is given by Eq.~(\ref{choice}).  The corresponding
squared mass in the propagator is given by
\begin{equation}
{m^2}'' = m^2 \e^{2 \Delta t} (1 + \ep)
\end{equation}
We have two adjustable parameters $\ep$, $\eta$ to satisfy two
vanishing conditions.  It is easy to check that with
\begin{eqnarray}
\eta &=& - \Delta t \frac{\partial}{\partial p^2} \left( \frac{1}{2}
\int_q \frac{\Delta (q)}{q^2} A_4 (q,-q,p,-p) \right)_{p^2=0}
\label{eta}\\ \ep + \eta &=& - \Delta t \frac{1}{2} \int_q \Delta (q)
\left( \frac{1}{q^2} B_4 (q,-q,0,0) - \frac{1}{q^4} A_4 (q,-q,0,0)
\right) \label{epsilon}
\end{eqnarray}
the modified vertices $\{\V''_{2n}\}$ satisfy the desired conditions:
\begin{equation}
\frac{d}{dp^2} A''_2 (p)\Big|_{p^2=0} = B''_2 (p=0) = 0
\end{equation}

Thus, we have obtained a modified ERG transformation which changes the
squared mass from $m^2$ to $m^2 \e^{2 \Delta t} (1 + \ep)$, and the
interaction vertices from $\{\V_{2n}\}$ to $\{\V''_{2n}\}$.  The
two conditions (\ref{conditions}) are preserved under the transformation.

We now wish to rewrite the above infinitesimal change as differential
equations.  For that purpose we first introduce a notation that makes
it clear that both $\ep$ and $\eta$ are proportional to $\Delta t$:
\begin{equation}
\ep = \beta_m \Delta t,\quad \eta = 2 \gamma \Delta t,
\label{defbetamgamma} 
\end{equation}
where $\beta_m$ is the anomalous dimension of the squared mass, and
$\gamma$ is that of the scalar field.  Denoting the $t$-dependence of
the squared mass and vertices by $m^2 (t)$ and $\{\V_{2n} (t;
p_1,\cdots,p_{2n})\}$, we obtain the following modified ERG
differential equations:
\begin{eqnarray}
&&\frac{d}{dt} m^2 (t) = \left( 2 + \beta_m (t) \right) m^2
(t),\label{diffm}\\ &&\frac{d}{dt} \left( \e^{- 2t} \V_2 (t; p \e^t)
\right) = \beta_m (t) m^2 (t) \e^{- 2t} + 2 \gamma (t) ( p^2 + m^2 (t)
\e^{-2t}) \nonumber\\ &&\quad + \left[ - 2 \gamma(t) + 2(1 - K(p
\e^t)) \left( \gamma (t) + \frac{\beta_m(t)}{2} \frac{m^2 (t)
\e^{-2t}}{p^2 + m^2 (t) \e^{-2t}} \right) \right] \e^{-2t} \V_2 (t; p
\e^t)\label{diffVtwo}\\ && \quad + \left(\e^{-2t} \V_2 (t; p \e^t)
\right)^2 \frac{\Delta (p \e^t)}{p^2 + m^2 (t) \e^{-2t}} + \frac{1}{2}
\int_q \frac{\Delta (q \e^t)}{q^2 + m^2 (t) \e^{-2t}} \V_4 (t; q \e^t,
-q \e^t, p \e^t, - p \e^t), \nonumber\\ &&\frac{d}{dt} \left( \e^{-
y_{2n}t} \V_{2n \ge 4} (t;p_1 \e^t,\cdots, p_{2n} \e^t) \right) =
\e^{- y_{2n} t} \V_{2n} (t; p_1,\cdots, p_{2n}) \nonumber\\ &&
\qquad\qquad\times \left[ - 2n \gamma(t) + \sum_{i=1}^{2n} (1 - K(p_i \e^t))
\left( \gamma (t) + \frac{\beta_m (t)}{2} \frac{m^2 (t)
\e^{-2t}}{p_i^2 + m^2 (t) \e^{-2t}} \right) \right] \nonumber\\ &&
\quad + \sum_{k=0}^{\left[\frac{n-1}{2}\right]}
\sum_{\text{partitions} \atop I+J=\{2n\}} \e^{- y_{2(k+1)}t}
\V_{2(k+1)} (t;p_I \e^t) \frac{\Delta (p_I \e^t)}{p_I^2 + m^2 (t)
\e^{-2t}} \e^{- y_{2(n-k)t}} \V_{2(n-k)} (t; p_J \e^t) \nonumber\\
&&\qquad\qquad + \frac{1}{2} \int_q \frac{\Delta (q \e^t)}{q^2 + m^2
(t) \e^{-2t}} \, \e^{- y_{2(n+1)}t} \V_{2(n+1)} (t; q \e^t, -q \e^t,
p_1 \e^t, \cdots, p_{2n} \e^t)
\label{diffVtwon} 
\end{eqnarray}
We should note that the Green functions change only multiplicatively
under the modified ERG transformation:
\begin{eqnarray}
&&\frac{\partial}{\partial t} \vev{ \phi (p_1 \e^t) \cdots \phi
(p_{2n-1} \e^t) \phi}_{m^2 (t); \V (t)}\nonumber\\ &=& (- 4n + y_{2n}
+ 2n \gamma (t)) \vev{ \phi (p_1 \e^t) \cdots \phi (p_{2n-1} \e^t)
\phi}_{m^2 (t); \V (t)}
\end{eqnarray}

To write down the above modified ERG differential equations, only the
anomalous dimensions $\beta_m$, $\gamma$ were necessary.  We did not
need any beta function of a coupling constant.  As we will see in the
next section, a beta function will be necessary when we convert the
above differential equations into integral equations.  To introduce a
beta function, we first define a running coupling constant by
\begin{equation}
- \lambda (t) \equiv A_4 (t; 0,0,0,0) \label{deflambda}
\end{equation}
Then, the modified ERG differential equation (\ref{diffVtwon}) for
$2n=4$ gives
\begin{equation}
\frac{d}{dt} ( - \lambda (t)) = 4 \gamma (t) \lambda (t) + \frac{1}{2}
\int_q \frac{\Delta (q)}{q^2} A_6 (t; q,-q,0,0,0,0)
\end{equation}
by taking $m^2 \to 0$ and $p_i \to 0$.  This can be written as
\begin{equation}
\frac{d}{dt} \lambda (t) = \beta (t) \label{difflambda}
\end{equation}
if we define the beta function by
\begin{equation}
\beta (t) \equiv - 4 \gamma (t) \lambda (t) - \frac{1}{2}
\int_q \frac{\Delta (q)}{q^2} A_6 (t; q,-q,0,0,0,0) \label{defbeta}
\end{equation}
For comparison, let us also write down the results of
(\ref{defbetamgamma}):
\begin{eqnarray}
2 \gamma (t) &\equiv& - \frac{1}{2} \frac{\partial}{\partial p^2}
\left( \int_q \frac{\Delta (q)}{q^2} A_4 (t; q,-q, p, -p)
\right)_{p^2=0} \label{defgamma}\\ \beta_m (t) + 2 \gamma (t) &\equiv&
\frac{1}{2} \int_q \Delta (q) \left( \frac{1}{q^4} A_4 (t;q,-q,0,0) -
\frac{1}{q^2} B_4 (t; q,-q,0,0) \right) \label{defbetam}
\end{eqnarray}
The relation of $\beta$, $\beta_m$, and $\gamma$ to the interaction
vertices is extremely simple.

Before concluding this section, we note that the modified differential
equations (\ref{diffm}, \ref{diffVtwo}, \ref{diffVtwon}) are valid not
only in the continuum limit but also for any theories in the theory
space.  Only in the continuum limit, though, the vertices are
completely determined by the squared mass $m^2$ and self-coupling
$\lambda$.  Hence, the beta function $\beta (t)$ and anomalous
dimensions $\beta_m (t), \gamma (t)$ become functions of the running
self-coupling $\lambda (t)$ alone.  We will show this in the next
section.  Outside the continuum limit, their $t$-dependence is not
solely given by $\lambda (t)$.

\section{\label{sec:integral}Integral equations and scaling}

We now wish to convert the modified ERG differential equations
(\ref{diffVtwo}, \ref{diffVtwon}) into integral equations, following
the procedure given in Ref.~\onlinecite{s02}.  The continuum limit of
the perturbative $\phi^4$ theory is characterized by the polynomial
(with respect to $t$) behavior of the vertices $\V_{2n} (t;
p_1,\cdots,p_{2n})$ as $t \to - \infty$.  Using this as a boundary
condition, we can integrate the ERG differential equations
(\ref{diffVtwo}, \ref{diffVtwon}) to obtain an infinite set of
integral equations.  As we have explained in Ref.~\onlinecite{s02} and
have summarized in sect.~\ref{sec:review}, the only undetermined
parameters of the integral equations are $m^2$, $B_2(0)$, $C_2(0)$,
and $A_4(0)$.  But in the present case, we have chosen $B_2(0) =
C_2(0) = 0$, and our integral equations have only two unknowns left,
namely the squared mass $m^2$ and the self-coupling constant $\lambda
= - A_4 (0)$.

We will not repeat the derivation of the integral equations since it
is explained in detail in Ref.~\onlinecite{s02}.  We only state the
results here.  Note that we have replaced the parameter $t$ by $-t$ so
that we will mainly deal with positive $t$ along an ERG trajectory.

For $2n \ge 6$, simple integration gives
\begin{eqnarray}
&&\e^{y_{2n} t} \V_{2n} (-t; p_1 \e^{-t},\cdots,p_{2n} \e^{-t})
\nonumber\\ &=& \int_0^\infty dt' \Bigg[ \, \e^{y_{2n} \tpt} \V_{2n}
(-\tpt; p_1 \e^{-\tpt},\cdots, p_{2n} \e^{-\tpt}) \Bigg\lbrace - 2n
\gamma(-\tpt) \nonumber\\ && \quad
 + \sum_{i=1}^{2n} (1 - K(p_i
\e^{-\tpt})) \left( \gamma (-\tpt) + \frac{\beta_m (-\tpt)}{2}
\frac{m^2 (-\tpt) \e^{2(t+t')}}{p_i^2 + m^2 (-\tpt) \e^{2(t+t')}}
\right) \Bigg\rbrace \nonumber\\ && +
\sum_{k=0}^{\left[\frac{n-1}{2}\right]} \sum_{\text{partitions}\atop
I+J=\{2n\}} \e^{-y_{2(k+1)}(t+t')} \V_{2(k+1)} (- \tpt; p_I \e^{-
\tpt}) \nonumber\\ &&\qquad\times \frac{\Delta (p_I \e^{-\tpt})}{p_I^2
+ m^2 (-\tpt) \e^{2(t+t')}} \e^{- y_{2(n-k)} (t+t')} \V_{2(n-k)}
(-\tpt; p_J \e^{-\tpt}) \label{inttwon}\\ && + \frac{1}{2} \int_q
\frac{\Delta (q \e^{-\tpt})}{q^2 + m^2 (t+t') \e^{2(t+t')}}
\nonumber\\ &&\qquad \times \e^{- y_{2(n+1)}(t+t')} \V_{2(n+1)}
(-\tpt; q \e^{-\tpt}, - q \e^{- (t+t')}, p_1 \e^{-\tpt},\cdots, p_{2n}
\e^{-\tpt}) \,\Bigg]\nonumber
\end{eqnarray} 
Thanks to $y_{2n} \le -2$, the integrand decreases at least as
$\e^{-2t'}$ for large $t'$, and the integral is automatically
convergent.  

For $2n=4$, we obtain
\begin{eqnarray}
&&\V_4 (-t; p_1 \e^{-t},\cdots, p_4 \e^{-t})\nonumber\\ &=&
\int_0^\infty dt' \Bigg[ \, \V_4 (-\tpt; p_1 \e^{-\tpt}, \cdots, p_4
\emi) \Bigg\lbrace - 4 \gamma(-\tpt) \nonumber\\ &&\: + \sum_{i=1}^4 (
1 - K(p_i \emi)) \left( \gamma (-\tpt) + \frac{\beta_m (-\tpt)}{2}
\frac{m^2 (-\tpt) \e^{2\tpt}}{p_i^2 + m^2 (-\tpt) \e^{2\tpt}} \right)
\Bigg\rbrace\nonumber\\ &&\quad + \sum_{i=1}^4 \e^{2 \tpt} \V_2
(-\tpt; p_i \emi) \frac{\Delta (p_i \emi)}{p_i^2 + m^2 (-\tpt)
\e^{2\tpt}}\nonumber\\ &&\qquad\qquad\qquad \times \V_4 (-\tpt; p_1
\emi, \cdots, p_4 \emi) \nonumber\\ &&\quad + \frac{1}{2} \int_q
\frac{\Delta (q \e^{-\tpt})}{q^2 + m^2 (-\tpt) \e^{2\tpt}} \nonumber\\
&&\qquad \times \e^{-2\tpt} \V_6 (-\tpt; q \emi, -q \emi, p_1 \emi,
\cdots, p_4 \emi) \nonumber\\ &&\quad + \beta (-\tpt) \quad
\Bigg]\quad - \lambda (-t)
\label{intfour} 
\end{eqnarray}
where the beta function $\beta (-(t+t'))$ is introduced to make the
integral convergent.  For large $t'$, the integrand behaves as
\begin{equation}
4 \gamma (-(t+t')) \lambda (-(t+t')) + \frac{1}{2} \int_q \frac{\Delta
(q)}{q^2} A_6 (-(t+t'); q,-q,0,0,0,0) + \beta (-(t+t')) + \text{O}
(\e^{-2t'})
\end{equation}
The polynomial terms cancel due to the definition of the beta function
(\ref{defbeta}), and this is of order $\e^{-2t'}$.  Hence, the
integral is finite.

Finally for $2n=2$ we obtain
\begin{eqnarray}
&&\e^{2t} \V_2 (-t; p \e^{-t})\nonumber\\
&=& \int_0^\infty dt' \Bigg[ \, - 2 \gamma (-\tpt) \e^{2(t+t')}
\left\lbrace \V_2 (-\tpt; p \emi) - A_2 (-\tpt) \right\rbrace
\nonumber\\
&&\quad + 2 (1- K(p \emi)) \left( \gamma (-\tpt) + \frac{\beta_m
(-\tpt)}{2} \frac{m^2 (-\tpt) \e^{2(t+t')}}{p^2 + m^2 (-\tpt)
\e^{2(t+t')}} \right)\nonumber\\
&&\qquad\qquad \times \, \e^{2(t+t')} \V_2 (-\tpt; p \emi) \nonumber\\
&&\quad + 2 \gamma (-\tpt) ( p^2 + m^2 (-\tpt) \e^{2(t+t')}) + \beta_m
(-\tpt) m^2 (-\tpt) \e^{2 (t+t')} \nonumber\\
&&\quad + \left( \e^{2(t+t')} \V_2 (-\tpt; p \emi) \right)^2
\frac{\Delta (p \emi)}{p^2 + m^2 (-\tpt) \e^{2(t+t')}} \nonumber\\
&&\:+ \frac{1}{2} \int_q \Bigg\{ \frac{\Delta (q \emi)}{q^2 + m^2
(-\tpt) \e^{2(t+t')}} \, \V_4 (-\tpt; q \emi, -q \emi, p \emi, - p
\emi)\nonumber\\ &&\qquad - \frac{\Delta (q \emi)}{q^2}\, A_4 (-\tpt;
q \emi, -q \emi, 0,0) \, \Bigg\}\, \Bigg]
\quad + \e^{2t} A_2 (-t) \label{inttwo}
\end{eqnarray}
To see the convergence of the integral, we use the conditions
\begin{equation}
\frac{\partial}{\partial p^2} A_2 (-t; p)\Bigg|_{p^2=0} = B_2 (-t) = 0
\label{tdepconditions}
\end{equation}
and find that for large $t'$ the integrand behaves as
\begin{eqnarray}
&&2 \gamma(-\tpt) (p^2 + m^2 (-\tpt)\e^{2\tpt}) + \beta_m (-\tpt) m^2
(-\tpt)\e^{2\tpt}\nonumber \\ &&\, + \frac{1}{2} p^2 \frac{\partial}{\partial
r^2} \left( \int_q \frac{\Delta (q)}{q^2} A_4 (-\tpt; q,-q,r,-r)
\right)_{r^2=0} \nonumber\\ &&\, + \frac{1}{2} m^2 (-\tpt)
\e^{2\tpt}\nonumber\\ &&\quad \times \int_q \Delta (q) \left(
\frac{B_4 (-\tpt; q,-q,0,0)}{q^2} - \frac{A_4 (-\tpt; q,-q,0,0)}{q^4}
\right) \: + \text{O} (\e^{-2t'})
\end{eqnarray}
Again, the polynomial terms cancel due to the definition of $\gamma$,
$\beta_m$ (\ref{defgamma}, \ref{defbetam}).

It is important to notice that the integral equation for $\V_2$
requires the knowledge of $A_2 (-t)$.  To determine $A_2 (-t)$ we must
go back to the modified ERG differential equation (\ref{diffVtwo}),
which gives
\begin{equation}
- \frac{d}{dt} \left( \e^{2t} A_2 (-t) \right) \nonumber\\ = \e^{2t}
\left[ - 2 \gamma (-t) A_2 (-t) + \frac{1}{2} \int_q \frac{\Delta
(q)}{q^2} \, A_4 (-t; q,-q,0,0) \right] \label{diffAtwo}
\end{equation}
Integrating this with respect to $t$, we obtain an integral equation
that determines $A_2 (-t)$ as
\begin{equation}
\e^{2t} A_2 (-t) = \int^t dt' \e^{2t'} \left[ 2 \gamma (-t') A_2 (-t')
- \frac{1}{2} \int_q \frac{\Delta (q)}{q^2}\, A_4 (-t'; q,-q,0,0)
\right]
\end{equation}
As it is, the right-hand side is ambiguous by a $t$-independent
constant.  We remove the ambiguity by introducing a convention that
\begin{equation}
\int^t dt' \e^{2t'} t'^k = \e^{2t} T_k (t)
\end{equation}
where $T_k (t)$ is an order $k$ polynomial of $t$.  This convention
gives $A_2 (-t)$ as a finite degree polynomial of $t$ at each order of
perturbation theory.  As we pointed out in Ref.~\onlinecite{s02}, this
convention amounts to the mass independence, i.e., $m^2=0$ corresponds
to the massless theory.

Although we did not derive the above integral equations, it is
straightforward to check that they reproduce the modified ERG
differential equations (\ref{diffVtwo}, \ref{diffVtwon}) by
differentiating them with respect to $t$.  It is also easy to check
that the integral equations give the correct asymptotic behavior as $t
\to \infty$:
\begin{eqnarray}
\e^{y_{2n} t} V_{2n} (-t; p_1 \e^{-t},\cdots, p_{2n} \e^{-t})
&\longrightarrow& 0 \quad \text{for} \quad 2n \ge 6\\
\V_4 (-t; p_1 \e^{-t},\cdots,p_4 \e^{-t}) &\longrightarrow& - \lambda
(-t)\\
\e^{2t} \V_2 (-t; p \e^{-t}) &\longrightarrow& \e^{2t} A_2 (-t) 
\end{eqnarray}
where the conditions (\ref{tdepconditions}) have been used.

To summarize, we have converted the differential equations
(\ref{diffVtwo}, \ref{diffVtwon}) into the integral equations
(\ref{inttwon}, \ref{intfour}, \ref{inttwo}) where $\beta$, $\beta_m$,
and $\gamma$ are defined by Eqs.~(\ref{defbeta}, \ref{defgamma},
\ref{defbetam}), $m^2 (-t)$ is the solution of Eq.~(\ref{diffm})
satisfying $m^2 (0) = m^2$, and $\lambda (-t)$ is 
defined by
\begin{equation}
- \lambda (-t) \equiv A_4 (-t; 0,0,0,0)
\end{equation}

As we have already mentioned in sect.~\ref{sec:intro}, the
subtractions necessary for the convergence of the integral equations
for the two- and four-point functions are determined by $\beta$,
$\beta_m$, and $\gamma$.  This is a nice property shared by the
minimal subtraction scheme in dimensional regularization, where all
the renormalization constants are determined by $\beta$, $\beta_m$,
$\gamma$.\cite{th73}

Our integral equations determine the vertices completely.  The only
necessary input is the squared mass $m^2$ and the self-coupling
$\lambda$ at the end ($t=0$) of the ERG trajectory $\{\V_{2n} (-t)\}$.
Once the input is given, all the vertices are determined unambiguously
in terms of the integral equations for the entire ERG trajectory
from $t=0$ to $\infty$.  To emphasize the necessity of the input
$m^2$, $\lambda$, we denote the solution of the integral equations by
\begin{equation}
\V_{2n} (-t; p_1,\cdots, p_{2n}; m^2, \lambda)
\end{equation}
and their coefficients of expansions in $m^2 (-t)$ by
\begin{equation}
A_{2n} (-t;p_1,\cdots,p_{2n};\lambda),\quad B_{2n} (-t;
p_1,\cdots,p_{2n}; \lambda)
\end{equation}

Now, let us make two crucial observations:
\begin{enumerate}
\item $\beta (-t)$, $\beta_m (-t)$, $\gamma (-t)$ are determined by
$\lambda$ and $t$, since these are determined by the vertices $A_6
(-t;q,-q,0,0,0,0; \lambda)$, $A_4 (-t; q,-q,p,-p; \lambda)$, $B_4 (-t;
q,-q,0,0;\lambda)$.
\item Given an ERG trajectory, we can take any point along the
trajectory as the origin $t=0$.  No matter where we start, we can
reproduce the entire trajectory by solving the integral equations.  In
other words, the following scaling relation holds:
\begin{equation}
\V_{2n} (-t-\Delta t; p_1,\cdots,p_{2n}; m^2, \lambda) = \V_{2n} (-t;
p_1,\cdots, p_{2n}; m^2 (-\Delta t), \lambda (-\Delta t)) \label{scaling}
\end{equation}
where $\Delta t$ is an arbitrary shift.  This is because the
conditions (\ref{tdepconditions}) are preserved under the modified ERG
transformation.
\end{enumerate}
\begin{figure}
\includegraphics{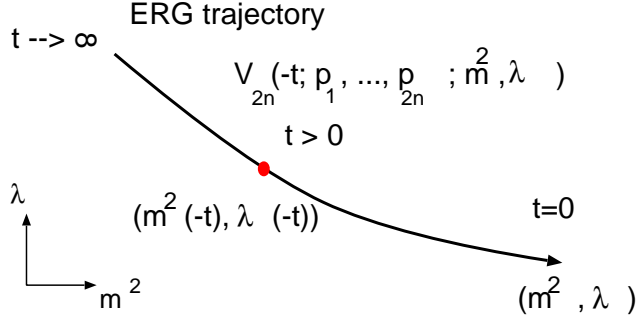}
\caption{ERG trajectory whose end point at $t=0$ is specified by
$(m^2, \lambda)$}
\end{figure}
Eq.~(\ref{scaling}) especially implies that 
\begin{equation}
\V_{2n} (-t; p_1,\cdots, p_{2n}; m^2, \lambda) = \V_{2n} (0;
p_1,\cdots,p_{2n}; m^2 (-t), \lambda (-t)) \label{altscaling}
\end{equation}
and hence
\begin{eqnarray}
A_6 (-t; q,-q,0,0,0,0; \lambda) &=& A_6 (0; q,-q, 0,0,0,0; \lambda
(-t))\\ A_4 (-t; q,-q, p,-p; \lambda) &=& A_4 (0; q,-q,p,-p; \lambda
(-t))\\ B_4 (-t; q,-q, 0,0; \lambda) &=& B_4 (0; q,-q,0,0; \lambda
(-t))
\end{eqnarray}
Therefore, using Eqs.~(\ref{defbeta}, \ref{defgamma}, \ref{defbetam}),
the above scaling relations imply that the $t$-dependence of the beta
function and anomalous dimensions is given by the running coupling
constant:
\begin{equation}
\gamma(-t) = \gamma (\lambda(-t)),\quad \beta_m (-t) = \beta_m
(\lambda (-t)),\quad \beta (-t)= \beta (\lambda (-t))
\end{equation}

Using the scaling relation (\ref{altscaling}), let us rewrite the
integral equations.  For simplicity, we introduce
\begin{equation}
f(p, m^2, \lambda) \equiv (1 - K(p)) \left(\gamma(\lambda) +
\frac{\beta_m (\lambda)}{2} \frac{m^2}{p^2 + m^2} \right)
\end{equation}
Then, omitting the first argument ``$t=0$'' of the vertices, we obtain
\begin{eqnarray}
&&\V_{2n \ge 6} (p_1,\cdots,p_{2n}; m^2, \lambda)\nonumber\\ &=&
\int_0^\infty dt\, \Bigg[ \, \e^{y_{2n} t} \V_{2n} (p_1 \e^{-t},
\cdots, p_{2n} \e^{-t}; m^2 (-t), \lambda (-t)) \nonumber\\
&&\qquad\qquad\qquad \times \left\{ - 2n \gamma (\lambda(-\tpt)) +
\sum_{i=1}^{2n} f(p_i \e^{-t}, m^2 (-t), \lambda(-t)) \right\}
\nonumber\\ &&\:+ \sum_{k=0}^{\left[ \frac{n-1}{2} \right]}
\sum_{\text{partitions}\atop I+J=\{2n\}} \e^{y_{2(k+1)}t} \V_{2(k+1)}
(p_I \e^{-t}; m^2 (-t), \lambda (-t)) \label{maintwon}\\
&&\qquad\qquad \times \frac{\Delta (p_I \e^{-t})}{p_I^2 + m^2 (-t)
\e^{2t}} \times \e^{y_{2(n-k)}t} \V_{2(n-k)} (p_J \e^{-t}; m^2 (-t),
\lambda (-t)) \nonumber\\ &&\: + \frac{1}{2} \int_q \frac{\Delta (q
\e^{-t})}{q^2 + m^2 (-t) \e^{2t}}\, \e^{y_{2(n+1)}t} \V_{2(n+1)} (q
\e^{-t}, -q \e^{-t}, p_1 \e^{-t}, \cdots, p_{2n} \e^{-t}; m^2 (-t),
\lambda (-t))\:\Bigg] \nonumber
\end{eqnarray}
For the four-point vertex we obtain
\begin{eqnarray}
&&\V_4 (p_1,\cdots,p_4; m^2, \lambda)\nonumber\\ &=& \int_0^\infty dt
\, \Bigg[ \: \V_4 (p_1 \e^{-t},\cdots, p_4 \e^{-t}; m^2 (-t), \lambda
(-t)) \nonumber\\ && \qquad\qquad\qquad \times \left\lbrace - 4
\gamma(\lambda(-t)) + \sum_{i=1}^4 f(p_i \e^{-t}, m^2 (-t),
\lambda(-t)) \right\rbrace\nonumber\\ &&\: + \frac{1}{2} \int_q
\frac{\Delta (q \e^{-t})}{q^2 + m^2 (-t) \e^{2t}} \, \e^{-2t} \V_6 (q
\e^{-t}, -q \e^{-t}, p_1 \e^{-t},\cdots, p_4 \e^{-t}; m^2 (-t),
\lambda (-t)) \nonumber\\ &&\: + \sum_{i=1}^4 \e^{2t} \V_2 (p_i
\e^{-t}; m^2 (-t), \lambda (-t)) \frac{\Delta (p_i \e^{-t})}{p_i^2 +
m^2 (-t) \e^{2t}} \V_4 (p_1 \e^{-t},\cdots, p_4 \e^{-t}; m^2 (-t),
\lambda (-t)) \nonumber\\ && \qquad\qquad + \quad \beta (\lambda
(-t))\quad \Bigg] - \lambda \label{mainfour}
\end{eqnarray}
Finally, for the two-point vertex we obtain
\begin{eqnarray}
&& \V_2 (p; m^2, \lambda) \nonumber\\ &=& \int_0^\infty dt \, \Bigg[
\, - 2 \gamma (\lambda (-t)) \, \e^{2t} \left\lbrace \V_2 (p \e^{-t};
m^2 (-t), \lambda (-t)) - a_2 (\lambda (-t)) \right\rbrace \nonumber\\
&&\quad + 2 f(p \e^{-t}, m^2 (-t), \lambda(-t)) \, \e^{2t} \V_2 (p
\e^{-t}; m^2 (-t), \lambda(-t))\nonumber\\ && + 2 \gamma
(\lambda(-t))\cdot ( p^2 + m^2 (-t) \e^{2t} ) + \beta_m (\lambda(-t))
\cdot m^2 (-t) \e^{2t} \nonumber\\ && + \frac{1}{2} \int_q \Bigg\{
\frac{\Delta (q \e^{-t})}{q^2 + m^2 (-t) \e^{2t}} \V_4 (q \e^{-t}, -q
\e^{-t}, p \e^{-t}, - p \e^{-t}; m^2 (-t), \lambda (-t)) \nonumber\\
&& \qquad\qquad - \frac{\Delta (q \e^{-t})}{q^2} A_4 (q \e^{-t}, -q
\e^{-t}, 0,0; \lambda (-t)) \Bigg\}\nonumber\\ && + \left( \e^{2t}
\V_2 (p \e^{-t}; m^2 (-t), \lambda (-t)) \right)^2 \frac{\Delta (p
\e^{-t})}{p^2 + m^2 (-t) \e^{2t}} \:\Bigg] + a_2 (\lambda)
\label{maintwo}
\end{eqnarray}
In the above $\lambda(-t)$ is defined by 
\begin{equation}
- \lambda (-t) \equiv A_4 (0,0,0,0; \lambda (-t)), \label{mainlambda}
\end{equation}
and it gives the solution of
\begin{equation}
- \frac{d}{dt} \lambda (-t) = \beta (\lambda (-t))
\end{equation}
satisfying the initial condition $\lambda (0) = \lambda$.  The running
squared mass is given by
\begin{equation}
m^2 (-t) \e^{2t} = m^2 \exp \left[ - \int_0^t dt' \beta_m (\lambda
(-t')) \right] \label{maindm}
\end{equation}
In the integral equation for $\V_2$, we have introduced a new notation
\begin{equation}
a_2 (\lambda) \equiv A_2 (0, 0; \lambda)
\end{equation}
Using the scaling relation, the differential equation (\ref{diffAtwo}) gives
\begin{equation}
2 a_2 (\lambda) = \beta (\lambda) a_2' (\lambda) + 2 \gamma (\lambda)
a_2 (\lambda) - \frac{1}{2} \int_q \frac{\Delta (q)}{q^2} A_4
(q,-q,0,0; \lambda) \label{mainatwo}
\end{equation}
which has a unique solution if $a_2 (\lambda)$ is obtained as a power
series in $\lambda$.

The integral equations (\ref{maintwon}, \ref{mainfour}, \ref{maintwo})
and the equations (\ref{mainlambda}, \ref{maindm}), and
(\ref{mainatwo}) constitute the main results of this section.  To
collect all the main results in a single place, we also repeat the
definitions (\ref{defbeta}, \ref{defgamma}, \ref{defbetam}) of the
beta function and anomalous dimensions, this time using the scaling
relation:
\begin{eqnarray}
2 \gamma (\lambda) &\equiv& - \frac{1}{2} \frac{\partial}{\partial
p^2} \left( \int_q \frac{\Delta (q)}{q^2}\, A_4 (q,-q,p,-p; \lambda)
\right)_{p^2=0} \label{scalegamma}\\ \beta_m (\lambda) + 2 \gamma
(\lambda) &\equiv& \frac{1}{2} \int_q \Delta (q) \left( \frac{A_4
(q,-q,0,0; \lambda)}{q^4} - \frac{B_4 (q,-q,0,0; \lambda)}{q^2}
\right)\label{scalebetam}\\ \beta (\lambda) + 4 \lambda \cdot \gamma
(\lambda) &\equiv& - \frac{1}{2} \int_q \frac{\Delta (q)}{q^2}\, A_6
(q,-q,0,0,0,0; \lambda) \label{scalebeta}
\end{eqnarray}

Our integral equations determine the continuum limit directly.  Solved
recursively, the integral equations give the interaction vertices as
power series in $\lambda$.  For a general algorithm, please refer to
sect.~IV of Ref.~\onlinecite{s02}.  The advantage of the integral
equations given above over those obtained in Ref.~\onlinecite{s02} is
twofold:
\begin{enumerate}
\item there are only two free parameters instead of four
\item the scaling relation (\ref{altscaling}) is valid for the
vertices
\end{enumerate}
The second point is especially meaningful: the entire scale dependence
of the interaction vertices reduces to that of the running parameters.
For the above two points, it was crucial that we have modified the ERG
transformation so that the conditions 
\begin{equation}
\frac{\partial}{\partial p^2} A_2 (p; \lambda) \Big|_{p^2=0} = B_2
(p=0; \lambda) = 0 \label{scaleconvention}
\end{equation}
are preserved.

The beta function $\beta$ and the anomalous dimensions $\beta_m,
\gamma$ depend on the choice of the momentum cutoff function $K(p)$.
We will discuss the dependence in the next section.  In Appendix
\ref{appendix:beta} we compute $\beta$, $\beta_m$, $\gamma$ up to
two-loop for the choice $K(p) = \theta (1-p^2)$, and show that the
non-universal part (the order $\lambda^2$ term of $\beta_m$) agrees
with the result of the minimal subtraction scheme in dimensional
regularization.

\section{\label{sec:linear}More general counterterms}

To prepare for the discussion of universality in sect.~\ref{sec:univ}
we need to generalize the construction of counterterms in
sect.~\ref{sec:counter}.

Our starting point is the modified action $S'[\phi]$ given by
Eq.~(\ref{Sprime}).  We introduce a further change of variables:
\begin{equation}
\phi (p) \longrightarrow \phi (p) \left( 1 + \frac{1}{2} ( K(p) t(p) +
u(p) ) \right)
\end{equation}
where we impose
\begin{equation}
t (p) = u(p) = 0 \quad \text{for} \quad p^2 < 1 \label{tuzero}
\end{equation}
This changes the action $S'[\phi]$ to the following:
\begin{eqnarray}
\tilde{S} [\phi] &=& \frac{1}{2} \int_p \phi (p) \phi (-p) \,
\frac{p^2 + m^2 + \ep m^2}{K(p)} \left( 1 + K(p) t(p) + u(p)
\right)\nonumber\\
&& - \sum_{n=1}^\infty \frac{1}{(2n)!} \int_{p_1,\cdots,p_{2n-1}} \phi
(p_1) \cdots \phi (p_{2n}) \nonumber\\
&&\qquad\qquad \times  \V'_{2n} (p_1,\cdots,p_{2n}) \left( 1 +
\frac{1}{2} \sum_{i=1}^{2n} (K(p_i)t(p_i) + u(p_i)) \right)
\end{eqnarray}
where the primed vertices $\V'_{2n}$ and $s(p)$ are given by
Eqs.~(\ref{Vtwoprime}, \ref{Vprime}) and (\ref{choice}), respectively.

The coefficient of the quadratic term can be calculated as follows:
\begin{eqnarray}
&& \left( \frac{p^2 + m^2 + \ep m^2}{K(p)} - \V'_2 (p) \right) \left(
1 + K(p) t(p) + u(p) \right) = \frac{p^2 + m^2 + \ep m^2}{K(p)} ( 1 +
u(p) )\nonumber\\ &&\qquad\qquad\qquad + t(p) (p^2 + m^2) - V'_2 (p) ( 1 + K(p)
t(p) + u(p) )
\end{eqnarray}
We treat the second line of the right-hand side as a vertex, and
define the propagator using only the first line.  The propagator is
then given by
\begin{equation}
\frac{K(p)}{p^2 + m^2 + \ep m^2} - \frac{K(p) u(p)}{p^2 + m^2}
\end{equation}
The condition (\ref{tuzero}) implies that the second term of the
propagator is non-vanishing only for $1 < p^2 < 2^2$.

\begin{figure}
\includegraphics{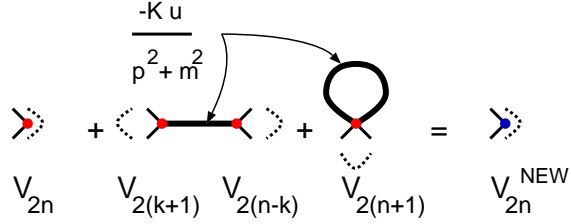}
\caption{\label{feynman}Reinterpretation of Feynman graphs gives new vertices.}
\end{figure}
In a Feynman diagram with the above propagator, we can interpret the
second term not as part of a propagator but as part of an interaction
vertex.  This reinterpretation is familiar from the derivation of the
exact renormalization group equation.\cite{pol84} (See
FIG.~\ref{feynman}.)  With this reinterpretation, the action
$\tilde{S}[\phi]$ is transformed into
\begin{eqnarray}
S''[\phi] &=& \int_p \frac{1}{2} \frac{p^2 + m^2 + \ep m^2}{K(p)} \, \phi
(p) \phi (-p) \nonumber\\
&&\quad - \sum_{n=1}^\infty \frac{1}{(2n)!}
\int_{p_1,\cdots,p_{2n-1}} \phi (p_1) \cdots \phi (p_{2n})\, \V''_{2n}
(p_1, \cdots, p_{2n})
\end{eqnarray}
where
\begin{eqnarray}
\V''_2 (p) &=& \ep m^2 + \eta (p^2 + m^2) - t(p) (p^2+m^2) + \V_2 (p)
( 1 + \tilde{s} (p))\nonumber\\ && - \V_2 (p) \frac{K(p) u(p)}{p^2 +
m^2} \V_2 (p) - \frac{1}{2} \int_q \frac{K(q) u(q)}{q^2 + m^2} \, \V_4
(q,-q,p,-p) \label{Vtwodoubleprime}\\ \V''_{2n} (p_1,\cdots,p_{2n})
&=& \V_{2n} (p_1,\cdots,p_{2n}) \left( 1 + \frac{1}{2} \sum_{i=1}^{2n}
\tilde{s}(p_i) \right) \nonumber\\ && -
\sum_{k=0}^{\left[\frac{n-1}{2}\right]} \sum_{\text{partitions}\atop
I+J=\{2n\}} \V_{2(k+1)} (p_I) \frac{K\left(p_I\right) u\left( p_I
\right)}{p_I^2 + m^2} \V_{2(n-k)} (p_J)\nonumber\\ && - \frac{1}{2}
\int_q \frac{K(q)u(q)}{q^2+m^2} \, \V_{2(n+1)}
(q,-q,p_1,\cdots,p_{2n}) \label{Vdoubleprime}
\end{eqnarray}
and
\begin{equation}
\tilde{s} (p) \equiv - \eta + (1 - K(p))
\left( \frac{\ep m^2}{p^2 + m^2} + \eta \right)  + K(p) t(p) + u(p)
\end{equation}

By construction the actions $S[\phi]$, given by (\ref{full}), and
$S''[\phi]$ are equivalent, since they are merely related by an
infinitesimal change of field variables
\begin{equation}
\phi (p) \longrightarrow \phi (p) \left( 1 + \frac{1}{2} \tilde{s} (p)
\right)
\end{equation}
From
\begin{equation}
\tilde{s} (p) = - \eta \quad \text{for}\quad p^2 < 1
\end{equation}
we obtain
\begin{equation}
\vev{ \phi (p_1) \cdots \phi (p_{2n-1}) \phi }_{m^2; \V}
= \left( 1 - n \eta \right) \vev{ \phi (p_1) \cdots \phi (p_{2n-1})
\phi }_{m^2 + \ep m^2; \V''} \label{generalequiv}
\end{equation}
if $(\forall i) p_i^2 < 1$.

Using only a linear change of field variables, the above counterterms
are as general as one can get, and this is just what we need in the
next section.  We will use the equivalence of $S[\phi]$ and
$S''[\phi]$ in the discussion of universality.

\section{\label{sec:univ}Universality}

Universality has been discussed in sect.~V of Ref.~\onlinecite{s02} in
the context of the original ERG equation.  The purpose of this section
is to provide a fuller discussion of universality for the modified ERG
equation, and thereby we wish to complete the arguments partially
presented in Ref.~\onlinecite{s02}.

The idea of universality has a broad meaning, but in this section we
restrict ourselves to the question of how the vertices in the
continuum limit depend on the choice of the momentum cutoff function
$K(p)$.  Given a choice of $K(p)$, all the interaction vertices $\{
V_{2n} \}$ are determined uniquely by a squared mass $m^2$ and
self-coupling constant $\lambda$, as we have seen in the previous
section.  When we change $K(p)$, the interaction vertices change even
with the same choice of $m^2$ and $\lambda$, and the Green functions
change accordingly.  In this section we wish to show that an arbitrary
infinitesimal change of $K(p)$ can be compensated by the corresponding
infinitesimal changes in $m^2$, $\lambda$, and normalization of the
field, and we wish to derive explicit formulas for the changes.

In sect.~\ref{sec:integral} we have concluded that the interaction
vertices are uniquely specified by a squared mass $m^2$, a
self-coupling constant $\lambda$, and a choice of the momentum cutoff
function $K (p)$.  To express the dependence explicitly, let us denote
the interaction vertices by $\{\V^K_{2n} (p_1,\cdots,p_{2n}; m^2,
\lambda)\}$.  We also denote the Green functions calculated with the
propagator
\begin{equation}
\frac{K(p)}{p^2+m^2} \label{originalprop}
\end{equation}
and the vertices $\{\V^K_{2n} (p_1,\cdots,p_{2n}; m^2,\lambda)\}$ by
\begin{equation}
\vev{ \phi (p_1) \cdots \phi (p_{2n-1}) \phi }^K_{m^2, \lambda}
\end{equation}
Using this notation, we can state what we wish to show in this section
more clearly.  We wish to show the existence of an infinitesimal
change $\ep (\lambda) m^2$ in the squared mass, $\delta \lambda
(\lambda)$ in the self-coupling, and $\eta (\lambda)$ in the
normalization of the scalar field so that
\begin{equation}
(1 - n \eta (\lambda)) \vev{ \phi (p_1) \cdots \phi (p_{2n-1}) \phi
}^{K+\delta K}_{m^2 (1+\ep),\lambda + \delta \lambda} = \vev{ \phi
(p_1) \cdots \phi (p_{2n-1}) \phi }^K_{m^2, \lambda}
\label{equivalence}
\end{equation}
where $\delta K (p)$ is an arbitrary infinitesimal change in the
momentum cutoff function satisfying \footnote{$\delta K(p)$ only needs
to be decreasing sufficiently fast for $p^2 > 2$.}
\begin{equation}
\delta K (p) = 0 \quad \text{unless} \quad 1 < p^2 < 2^2
\end{equation}

We first observe that the left-hand side of Eq.~(\ref{equivalence})
can be calculated using the propagator
\begin{equation}
\frac{K(p)}{p^2 + m^2 (1+\ep)}
\end{equation}
and the following vertices:
\begin{eqnarray}
&&\overline{\V}_{2n} (p_1,\cdots,p_{2n}; m^2 (1+\ep), \lambda + \delta
\lambda) \equiv \V_{2n}^{K+\delta K} (p_1, \cdots, p_{2n}; m^2
(1+\ep), \lambda + \delta \lambda)\nonumber\\ &&\quad +
\sum_{k=0}^{\left[\frac{n-1}{2}\right]} \sum_{\text{partitions}\atop I
+ J = \{2n\}} \V_{2(k+1)}^K (p_I; m^2, \lambda) \frac{\delta
K(p_I)}{p_I^2 + m^2} \V_{2(n-k)}^K (p_J; m^2, \lambda)\nonumber\\
&&\quad + \frac{1}{2} \int_q \frac{\delta K(q)}{q^2 + m^2}\,
\V_{2(n+1)}^K (q,-q,p_1,\cdots,p_{2n}; m^2, \lambda) \label{Vbar}
\end{eqnarray}
This results from the same reasoning as given in the previous section:
we reinterpret $\delta K$ not as part of a propagator but as part of
an interaction vertex.  Now, the equivalence (\ref{equivalence}) can
be reexpressed as
\begin{equation}
(1 - n \eta (\lambda)) \vev{ \phi (p_1) \cdots \phi (p_{2n-1}) \phi
}_{m^2 (1+\ep); \overline{\V}} = \vev{ \phi (p_1) \cdots \phi
(p_{2n-1}) \phi }_{m^2; \V} \label{intermediate}
\end{equation}
where we use the same momentum cutoff function $K(p)$ for both sides.

To go further, we need the results obtained in the previous section.
Using Eq.~(\ref{generalequiv}), we can rewrite the right-hand side of
Eq.~(\ref{intermediate}) to obtain
\begin{equation}
\vev{ \phi (p_1) \cdots \phi (p_{2n-1}) \phi}_{m^2 (1 + \ep);
\overline{\V}} = \vev{ \phi (p_1) \cdots \phi (p_{2n-1}) \phi }_{m^2
(1 + \ep); \V''} \label{altequivalence}
\end{equation}
where the vertices $\{\V''_{2n}\}$s are given by
(\ref{Vtwodoubleprime}, \ref{Vdoubleprime}) with the same $\ep$ as for
the left-hand side.  This equality is of course valid if the two sets
of vertices $\{\overline{\V}_{2n}\}$ and $\{\V''_{2n}\}$ are equal.

The original problem was to show the existence of $\delta
\lambda(\lambda)$, $\ep (\lambda)$, and $\eta (\lambda)$ so that
Eq.~(\ref{equivalence}) holds.  Now the problem is reduced to showing
the existence of $\delta \lambda (\lambda)$, $\ep (\lambda)$, $\eta
(\lambda)$, and functions $t(p)$, $u(p)$ so that
$\{\overline{\V}_{2n}\}$ and $\{\V''_{2n}\}$ become equal.  In other
words, given an arbitrary infinitesimal $\delta K(p)$, we wish to
determine $\delta \lambda (\lambda)$, $\ep (\lambda)$, $\eta
(\lambda)$, $u(p)$, and $t (p)$ so that the following equalities hold:
\begin{eqnarray}
&& \V^{K+\delta K}_2 (p; m^2 (1+\ep (\lambda)), \lambda + \delta
\lambda (\lambda))\nonumber\\ &=& \ep (\lambda) m^2 + \eta (\lambda)
(p^2 + m^2) - t(p) (p^2+m^2) + \V_2 (p; m^2,\lambda) ( 1 + \tilde{s}
(p))\nonumber\\ && \quad - \V_2 (p; m^2,\lambda) \frac{\delta K(p) +
K(p) u(p)}{p^2 + m^2} \V_2 (p; m^2, \lambda) \nonumber\\ &&\quad -
\frac{1}{2} \int_q \frac{\delta K(q) + K(q) u(q)}{q^2 + m^2} \, \V_4
(q,-q,p,-p; m^2, \lambda) \label{toshowtwo}
\end{eqnarray}
and
\begin{eqnarray}
&&\V^{K+\delta K}_{2n \ge 4} (p_1,\cdots,p_{2n}; m^2 (1 + \ep
(\lambda)), \lambda + \delta \lambda (\lambda))\nonumber\\ &=& \V_{2n}
(p_1,\cdots,p_{2n}; m^2, \lambda) \left( 1 + \frac{1}{2}
\sum_{i=1}^{2n} \tilde{s}(p_i) \right) \nonumber\\ && \quad -
\sum_{k=0}^{\left[\frac{n-1}{2}\right]} \sum_{\text{partitions}\atop
I+J=\{2n\}} \V_{2(k+1)} (p_I;m^2,\lambda) \frac{\delta K(p_I) +
K\left(p_I\right) u\left( p_I \right)}{p_I^2 + m^2} \V_{2(n-k)} (p_J;
m^2, \lambda)\nonumber\\ && \quad - \frac{1}{2} \int_q \frac{\delta
K(q) + K(q)u(q)}{q^2+m^2} \, \V_{2(n+1)} (q,-q,p_1,\cdots,p_{2n}; m^2,
\lambda) \label{toshowtwon}
\end{eqnarray}
where
\begin{equation}
\tilde{s} (p) \equiv - \eta(\lambda) + (1 - K(p)) \left( \frac{\ep(\lambda)
m^2}{p^2 + m^2} + \eta (\lambda) \right) + K(p) t(p) + u(p)
\end{equation}
Note that we have omitted the superscript $K$ from the vertices on the
right-hand sides.  The functions $t(p)$, $u(p)$ depend also on
$\lambda$ and $m^2$, and more appropriate notations would be $t(p;
m^2, \lambda)$ and $u(p; m^2, \lambda)$.

In the remainder of this section, we will show that with an
appropriate choice for the functions $u(p; m^2, \lambda)$ and $t(p;
m^2, \lambda)$, the right-hand sides of Eqs.~(\ref{toshowtwo},
\ref{toshowtwon}) satisfy the ERG equations expected of the left-hand
sides.  Since the vertices are determined completely by the ERG
equations, this will prove the equality.

The proof is given in three steps.  In the first step, we notice that
$\ep (\lambda)$, $\eta (\lambda)$, and $\delta \lambda (\lambda)$ are
determined by $\delta K(p)$ and $u(p)$ as a consequence of the
convention (\ref{scaleconvention}) and the definition
(\ref{mainlambda}) of $\lambda$.  Using the obvious notation, and
assuming Eqs.~(\ref{toshowtwo}, \ref{toshowtwon}) are valid, we find
the following:
\begin{eqnarray}
&& \frac{\partial}{\partial p^2} A_2^{K+\delta K} (p; \lambda + \delta
\lambda) \Big|_{p^2=0}\nonumber\\ &&\qquad = \eta - \frac{1}{2}
\frac{\partial}{\partial p^2} \left( \int_q \frac{\delta
K(q)+K(q)u(q;0,\lambda)}{q^2} \, A_4 (q,-q,p,-p) \right)_{p^2=0} \\ &&
B_2^{K+\delta K} (0;\lambda + \delta \lambda)\nonumber\\ &&\qquad =
\ep + \eta - \frac{1}{2} \int_q (\delta K(q) + K(q)u(q;0,\lambda))
\left( - \frac{A_4 (q,-q,0,0)}{q^4} + \frac{B_4 (q,-q,0,0)}{q^2}
\right) \nonumber\\ &&\qquad\qquad - \frac{1}{2} \int_q \frac{K(q)
\frac{\partial}{\partial m^2} u(q; m^2,\lambda)\Big|_{m^2=0}}{q^2} A_4
(q,-q,0,0) \\ &&A_4^{K+\delta K} (0,0,0,0;\lambda + \delta \lambda)
\nonumber\\ &&\qquad = (-\lambda) (1 - 2 \eta) - \frac{1}{2} \int_q
\frac{\delta K(q)+K(q)u(q;0,\lambda)}{q^2} \, A_6 (q,-q,0,0,0,0)
\end{eqnarray}
For Eqs.~(\ref{toshowtwo}, \ref{toshowtwon}) to be valid, these three
equations must be satisfied.  The first two must vanish, while the
last must equal $-(\lambda + \delta \lambda)$.  Hence, we must obtain
\begin{eqnarray}
\eta &=& \frac{1}{2} \frac{\partial}{\partial p^2} \left( \int_q
\frac{\delta K(q)+ K(q)u(q;0,\lambda)}{q^2} \, A_4 (q,-q,p,-p)
\right)_{p^2=0}
\label{etanew} \\ \ep +
\eta &=& \frac{1}{2} \int_q (\delta K(q) + K(q) u(q;0,\lambda)) \left(
- \frac{1}{q^4} A_4 (q,-q,0,0) + \frac{1}{q^2} B_4 (q,-q,0,0)
\right)\nonumber\\ &&\qquad + \frac{1}{2} \int_q \frac{K(q)
\frac{\partial}{\partial m^2} u(q;m^2,\lambda)\Big|_{m^2=0}}{q^2}\,
A_4 (q,-q,0,0) \label{epsilonnew} \\ \delta \lambda + 2 \lambda \cdot
\eta &=& \frac{1}{2} \int_q \frac{\delta K(q) + K(q)
u(q;0,\lambda)}{q^2} \, A_6 (q,-q,0,0,0,0) \label{deltalambda}
\end{eqnarray}
Thus, $\ep$, $\eta$, and $\delta \lambda$ are determined by $\delta K$
and $u$.  Therefore, we only need to determine the functions
$u(p;m^2,\lambda)$ and $t(p; m^2, \lambda)$.

The second step is to determine the changes in the beta function and
anomalous dimensions due to the changes of the parameters $m^2,
\lambda$ and normalization of the field.  The result is well known.
By considering the running of $\lambda + \delta \lambda$, $m^2
(1+\ep)$, and $(1 + \eta/2) \phi$, we obtain the following results up
to first order in $\delta K$:
\begin{eqnarray}
\beta^{K+\delta K} (\lambda + \delta \lambda) - \beta (\lambda) 
&=& \delta \lambda' (\lambda) \cdot \beta (\lambda)\\
\beta_m^{K+\delta K} (\lambda + \delta \lambda) - \beta_m (\lambda)
&=& \ep' (\lambda) \cdot \beta (\lambda)\\
\gamma^{K+\delta K} (\lambda + \delta \lambda) - \gamma (\lambda) &=&
\frac{1}{2} \eta' (\lambda) \cdot \beta (\lambda)
\end{eqnarray}
where the primes denote derivatives with respect to $\lambda$.  We do
not actually need $\beta^{K+\delta K}$ since only $\beta_m^{K+\delta
K}$ and $\gamma^{K+\delta K}$ enter into the modified ERG differential
equations for $\{\V^{K+\delta K}_{2n}\}$.

In the final step we demand that the right-hand sides of
(\ref{toshowtwo}, \ref{toshowtwon}) satisfy the modified ERG
differential equations expected of $\{\V^{K+\delta K}_{2n} (m^2
(1+\ep),\lambda + \delta \lambda)\}$.  This step is straightforward,
but requires somewhat lengthy calculations.  We omit the intermediate
results.  Remarkably, all boil down to the following equations for
$u(p;m^2, \lambda)$ and $t(p;m^2, \lambda)$:
\begin{eqnarray}
\frac{d}{dt} u(p) &=& (\delta K(p) + K(p) u(p)) \left( 2 \gamma +
\beta_m \frac{m^2}{p^2+m^2}\right) + \Delta (p) \left( \frac{\ep
m^2}{p^2+m^2} + \eta - t(p) \right) \label{dudt} \\
\frac{d}{dt} t(p) &=& u(p) \left( 2 \gamma + \beta_m \frac{m^2}{p^2 +
m^2} \right)\label{dtdt}
\end{eqnarray}
where we have defined the RG differential operator acting on functions
of $p^2$, $m^2$, and $\lambda$ as
\begin{equation}
\frac{d}{dt} \equiv 2 p^2 \frac{\partial}{\partial p^2} + \beta
(\lambda) \frac{\partial}{\partial \lambda} + (2 + \beta_m (\lambda))
m^2 \frac{\partial}{\partial m^2}
\end{equation}
The above differential equations can be solved perturbatively in
powers of $\lambda$, and thanks to the vanishing condition
(\ref{tuzero}), the solutions for $u(p), t(p)$ are unique.  It is easy
to see that $u$ is of order $\lambda$, and $t$ is of order
$\lambda^2$.  We will compute $u(p)$ and $t(p)$ to the lowest
nontrivial order in $\lambda$ in Appendix \ref{appendix:lowest}.  This
concludes the proof of the equality $\overline{\V} = \V''$, which
proves Eq.~(\ref{equivalence}).  Thus, we have established
universality of the continuum limit.

Finally, a short comment is in order, regarding sect.~V of
Ref.~\onlinecite{s02}.  If we demand that $\V^{K+\delta K}$ given by
(\ref{toshowtwo}, \ref{toshowtwon}) satisfy the original ERG equation
with no $\beta, \beta_m$, or $\gamma$, we obtain
\begin{equation}
u(p) = (1 - K(p)) \left( \frac{\ep m^2}{p^2 + m^2} + \eta \right)
\end{equation}
and $t(p)=0$, where $\ep, \eta$ are arbitrary infinitesimal constants.
This is easily obtained from Eqs.~(\ref{dudt}, \ref{dtdt}) by taking
$\beta_m$ and $\gamma$ to zero.  This result was quoted without proof
in sect.~V of Ref.~\onlinecite{s02}.  This will be also used in
Appendix \ref{appendix:improvement}.

\section{\label{sec:conclusion}Concluding remarks}

It is easy to summarize what we have done in this paper.  We have
united Wilson's exact renormalization group with the standard
renormalization group of running parameters in renormalized field
theories.  We have modified the ERG differential equation in such a
way that the continuum limit is parameterized by a squared mass and a
self-coupling constant.  The scale dependence of the infinite number
of interaction vertices is given through the scale dependence of the
two running parameters.  We have called this a scaling relation, and
it is expressed by Eq.~(\ref{scaling}) or (\ref{altscaling}).

The integral equation that we have introduced in Ref.~\onlinecite{s02}
and the present paper can be a powerful tool.  It is a defining
equation of a so-called ``perfect action'' that describes the
continuum limit in terms of a theory with a finite momentum cutoff.
It is especially interesting to see how the symmetry of the continuum
limit, such as chiral symmetry and gauge symmetry, is realized in a
perfect action.  We believe that our integral equation will provide a
powerful quantitative tool to address this question.

As one of the nice properties of the framework introduced in this
paper, we have mentioned the mass independence in
sect.~\ref{sec:intro}.  The mass independence means not only that the
squared mass renormalizes multiplicatively, but also that we have a
broken phase for $m^2 < 0$.  We have discussed the broken phase in
Appendix \ref{appendix:SSB}.

Concerning the issue of universality discussed in
sect.~\ref{sec:univ}, we have one technical remark.  We know that the
beta function and anomalous dimensions depend on the choice of the
momentum cutoff function $K(p)$.  Now, what choice gives the same
$\beta, \beta_m, \gamma$ as the minimal subtraction scheme in
dimensional regularization?  We speculate the step function $K(p) =
\theta (1 - p^2)$ is the answer, but this is only supported by a
single calculation of $\beta_m$ at order $\lambda^2$ in Appendix
\ref{appendix:beta}.  It will be interesting to know the answer.

In calculating the vertices recursively using the integral equations,
we have noticed the great facility brought by expressing the integrand
(with respect to $t$) as a total derivative.  Following the recursive
procedure blindly is a tedious task.  It will be extremely interesting
and useful to come up with a short-cut procedure to construct the
vertices in the continuum limit.

\appendix

\section{\label{appendix:beta}Explicit calculations of $\beta$, $\beta_m$, and
$\gamma$}

In this appendix we sketch the calculations of the beta function and
anomalous dimensions.  The style of presentation is not uniform: some
parts are given in more detail than others.  As is usual with
perturbative calculations, the difficulty is mainly in doing momentum
integrals.

We first recall the general formulas for the beta function $\beta
(\lambda)$ and anomalous dimensions $\beta_m (\lambda)$, $\gamma
(\lambda)$:
\begin{eqnarray}
2 \gamma (\lambda) &\equiv& - \frac{\partial}{\partial p^2} \left[
\frac{1}{2} \int_q \frac{\Delta (q)}{q^2} \, A_4 (q,-q,p,-p; \lambda)
\right]_{p^2=0} \\
2 \gamma (\lambda) + \beta_m (\lambda) &\equiv& \frac{1}{2} \int_q
\left\lbrace \frac{\Delta (q)}{q^4} A_4 (q,-q,0,0;\lambda) -
\frac{\Delta (q)}{q^2} B_4 (q,-q,0,0;\lambda) \right\rbrace\\
\beta (\lambda) + 4 \lambda \cdot \gamma (\lambda) &\equiv& - \frac{1}{2}
\int_q \frac{\Delta (q)}{q^2} A_6 (q,-q,0,0,0,0; \lambda)
\end{eqnarray}
where $A_4, B_4$, and $A_6$ are the coefficients of the vertices expanded in
powers of $m^2$:
\begin{equation}
\V_{2n} (p_1, \cdots, p_{2n}; m^2, \lambda) = A_{2n} (p_1, \cdots,
p_{2n}; \lambda) + m^2 B_{2n} (p_1,\cdots, p_{2n}; \lambda) + \text{O}
(m^4)
\end{equation}
$\beta$ is universal to order $\lambda^3$, $\gamma$ to order
$\lambda^2$, and $\beta_m$ only up to order $\lambda$.  The
non-universal higher order terms depend on the choice of the momentum
cutoff function $K(p)$.  In the following we evaluate $\beta_m,
\gamma$ to order $\lambda^2$, and $\beta$ to order $\lambda^3$.  Only
the order $\lambda^2$ term of $\beta_m$ is non-universal.  Except for
this non-universal term and the third order term of $\beta$, the
lowest order terms have been computed in Ref.~\onlinecite{hl88} by
Hughes and Liu.  Our results must and do agree with theirs because of
the universality.

We will need to calculate $A_4, B_4$ to order $\lambda^2$, and $A_6$
to order $\lambda^3$.  We will use the notation such as
\begin{equation}
A_2 (p; \lambda) = \sum_{i=1}^\infty (- \lambda)^i A_2^{(i)} (p),\quad
\V_2 (p; \lambda,m^2) = \sum_{i=1}^\infty (- \lambda)^i \V_2^{(i)} (p; m^2)
\end{equation}
to denote the order of expansions in $\lambda$.  We refer the reader
to section IV of Ref.~\onlinecite{s02} for a general algorithm for
perturbative calculations of the interaction vertices.

\subsection{Order $\lambda$}

\subsubsection{$4$-point}

Our starting point is
\begin{equation}
\V_4^{(1)} (p_1,\cdots,p_4) = 1
\end{equation}
implying 
\begin{equation}
A_4 ^{(1)} (p_1,\cdots,p_4) = 1,\quad B_4^{(1)} (p_1,\cdots,p_4) = 0
\end{equation}
Hence,
\begin{equation}
\gamma^{(1)} = 0
\end{equation}
and
\begin{equation}
\beta_m^{(1)} = \frac{1}{2} \int_q \frac{\Delta (q)}{q^4} A_4^{(1)} =
\frac{1}{2} \int_q \frac{\Delta (q)}{q^4} = \frac{1}{(4\pi)^2}
\label{betamfirst} 
\end{equation}
where the integrand is a total derivative due to Eq.~(\ref{Delta}).
Hence, the running squared mass is given by
\begin{equation}
m^2 (-t) \e^{2t} = m^2 \left( 1 - t \cdot (-\lambda) \cdot \beta_m^{(1)} +
\text{O} (\lambda^2) \right)
\end{equation}

\subsubsection{$2$-point}

Eq.~(\ref{mainatwo}) gives
\begin{equation}
a_2^{(1)} = - \frac{1}{4} \int_q \frac{\Delta (q)}{q^2} A_4^{(1)} = 
 - \frac{1}{4} \int_q \frac{\Delta (q)}{q^2}
\end{equation}
This depends on the choice of the cutoff function $K(p)$.  We then
find
\begin{eqnarray}
\V_2^{(1)} (m^2) &=& \int_0^\infty dt \left[ \frac{1}{2} \int_q
\left\{ \frac{\Delta (q \e^{-t})}{q^2 + m^2} - \frac{\Delta( q
\e^{-t})}{q^2} \right\} + \beta_m^{(1)} m^2 \right] + a_2^{(1)}
\nonumber\\
&=& \frac{m^4}{2} \int_q \frac{1-K(q)}{q^4 (q^2+m^2)} + a_2^{(1)}
\end{eqnarray}
Note that the two-point vertex is momentum independent at this order.

\subsection{Order $\lambda^2$}

\subsubsection{$6$-point}

From
\begin{equation}
\V_6^{(2)} (p_1,\cdots,p_6; m^2) = \frac{1 -
K(p_1+p_2+p_3)}{(p_1+p_2+p_3)^2 + m^2} + \text{9 permutations}
\end{equation}
we obtain
\begin{equation}
A_6^{(2)} (q,-q,0,0,0,0) = 6 \frac{1-K(q)}{q^2}
\end{equation}
Hence, we get
\begin{equation}
\beta^{(2)} = - \frac{1}{2} \int_q \frac{\Delta (q)}{q^2} A_6^{(2)}
(q,-q,0,0,0,0) = - 3 \int_q \frac{\Delta (q) (1 - K(q))}{q^4} = -
\frac{3}{(4\pi)^2} 
\end{equation}
using
\begin{equation}
\Delta (q) (1 - K(q)) = q^2 \frac{d}{dq^2} (1-K(q))^2
\end{equation}
Therefore, the running coupling is given by
\begin{equation}
\lambda (-t) = \lambda - t (-\lambda)^2 \beta^{(2)} + \text{O}
(\lambda^3)
\end{equation}

\subsubsection{$4$-point}

The integral ERG equation (\ref{mainfour}) gives
\begin{eqnarray}
&& \V_4^{(2)} (p_1,\cdots,p_4; m^2)\nonumber\\
&=& \int_0^\infty dt \, \Bigg[ \,
\sum_{i=1}^4 \e^{2t} \V_2^{(1)} (m^2 \e^{-2t}) \frac{\Delta(p_i
\e^{-t})}{p_i^2 + m^2}  + \frac{1}{2} \beta_m^{(1)} m^2 
\sum_{i=1}^4 \frac{1 - K(p_i \e^{-t})}{p_i^2 + m^2} \nonumber\\
&& \qquad + \frac{1}{2} \int_q \frac{\Delta (q \e^{-t})}{q^2 + m^2}
\Bigg\lbrace \sum_{i=1}^4 \frac{1 - K(p_i \e^{-t})}{p_i^2 + m^2}
\nonumber\\
&& \qquad \qquad + 2 \left( \frac{1 -
K((p_1+p_2+q)\e^{-t})}{(p_1+p_2+q)^2 + m^2} + \text{2 permutations}
\right) \Bigg\rbrace \, + \beta^{(2)} \: \Bigg]
\end{eqnarray}
Using the ERG differential equation
\begin{equation}
- \frac{\partial}{\partial t} \left( \e^{2t} \V_2^{(1)} (m^2 \e^{-2t})
\right) = \frac{1}{2} \int_q \frac{\Delta (q \e^{-t})}{q^2 + m^2} +
m^2 \beta_m^{(1)}
\end{equation}
and integration by parts, we obtain
\begin{eqnarray}
&&\V_4^{(2)} (p_1,\cdots,p_4; m^2)
= \V_2^{(1)} (m^2) \sum_{i=1}^4 \frac{1 - K(p_i)}{p_i^2 +
m^2} - \frac{1}{2} \beta_m^{(1)} m^2 \sum_{i=1}^4 \frac{\int_0^\infty dt
(1 - K(p_i \e^{-t}))}{p_i^2 + m^2}\nonumber\\
&&\quad + \int_0^\infty dt \left[ \, \int_q \frac{\Delta (q \e^{-t})}{q^2 +
m^2} \left( \frac{1 - K((p_1+p_2+q) \e^{-t})}{(p_1+p_2+q)^2 + m^2} +
\text{2 permutations} \right) + \beta^{(2)} \: \right]
\end{eqnarray}
Expanding this in powers of $m^2$, we obtain
\begin{eqnarray}
&& A_4^{(2)} (p,-p,0,0) = 2 a_2^{(1)} \frac{1 - K(p)}{p^2} \nonumber\\
&& \qquad\qquad+ 2 \int_0^\infty dt \int_q \frac{\Delta(q
\e^{-t})}{q^2} \left( \frac{1 - K((p+q)\e^{-t})}{(p+q)^2} -
\frac{1-K(q \e^{-t})}{q^2} \right)\\&& B_4^{(2)} (p,-p,0,0) = - 2
a_2^{(1)} \frac{1-K(p)}{p^4} - \beta_m^{(1)} \frac{\int_0^\infty dt (1
- K(p \e^{-t}))}{p^2} \nonumber\\ &&\qquad\qquad - 2 \int_q \frac{1 -
K(q)}{q^4} \frac{1 - K(p+q)}{(p+q)^2} - \int_q \frac{\Delta
(q)(1-K(q))}{q^6}
\end{eqnarray} 
and for $p'^2 < 1$ we obtain
\begin{eqnarray}
&&A_4^{(2)} (p',-p',p,-p) = 2 a_2^{(1)} \frac{1 - K(p)}{p^2}
\nonumber\\
&& + \int_0^\infty dt \, \Bigg[ \, \int_q \frac{\Delta (q
\e^{-t})}{q^2} \left( \frac{1 - K((p+p'+q)\e^{-t})}{(p+p'+q)^2} +
\frac{1 - K((p-p'+q)\e^{-t})}{(p-p'+q)^2} \right)\nonumber\\
&& \qquad \qquad - 2 \int_q \frac{\Delta (q) (1 - K(q))}{q^4} \, \Bigg]
\end{eqnarray}

We first compute
\begin{eqnarray}
- 2 \gamma^{(2)} &=& \frac{1}{2} \frac{\partial}{\partial p'^2} \int_p
  \frac{\Delta (p)}{p^2} A_4^{(2)} (p',-p',p,-p) \Bigg|_{p'^2 =
  0}\nonumber\\
&=& \int_p \frac{\Delta (p)}{p^2} \int_0^\infty dt \int_q \frac{\Delta
  (q \e^{-t})}{q^2} \frac{\partial}{\partial p'^2} \frac{1 -
  K((p+q+p')\e^{-t})}{(p+q+p')^2} \Bigg|_{p'^2 = 0}\nonumber\\
&=& - \frac{1}{2} \int_0^\infty dt \int_p \frac{\Delta (p \e^t)}{p^2}
  \int_q \frac{\Delta (q)}{q^2} K'' (p+q)\nonumber\\
&=& \int_{p,q} \frac{K(p)}{p^2} K' (q) K'' (p+q) \label{gamma}
\end{eqnarray}
where we used
\begin{equation}
\Delta (p \e^t) = - \partial_t K(p \e^t)\, , \label{dtK}
\end{equation}
and the primed quantities are defined by
\begin{equation}
K'(p) \equiv \frac{d}{dp^2} K(p),\quad K'' (p) \equiv \frac{d^2}{(d
p^2)^2} K(p)
\end{equation}
The last expression in Eq.~(\ref{gamma}) was calculated by Hughes and
Liu \cite{hl88} as 
\begin{equation}
- 2 \gamma^{(2)} = - \frac{1}{6} \frac{1}{(4\pi)^2} \label{gammaresult}
\end{equation}

We next compute
\begin{eqnarray}
\beta_m^{(2)} + 2 \gamma^{(2)} &=& \frac{1}{2} \int_p \left\{
\frac{\Delta (p)}{p^4} A_4^{(2)} (p,-p,0,0) - \frac{\Delta (p)}{p^2}
B_4^{(2)} (p,-p,0,0) \right\}\nonumber\\
&=& \frac{1}{4} \int_q \frac{\Delta (q)}{q^4} \int_p
\frac{(1-K(p))K(p)}{p^4} \nonumber\\
&&+ \int_p \frac{K(p)}{p^4} \int_q \frac{1-K(q)}{q^2} \left\{
\frac{\Delta (p+q)}{(p+q)^2} - \frac{\Delta(q)}{q^2}
\right\}\nonumber\\
&&+ \int_p \frac{K(p)}{p^2} \int_q \left\{ \frac{1-K(q)}{q^2}
\frac{\Delta (p+q)}{(p+q)^4} + \frac{1-K(q)}{q^4} \frac{\Delta
(p+q)}{(p+q)^2} \right\} \label{betamgamma}
\end{eqnarray}
where we have used Eq.~(\ref{dtK}) to perform several integrals over
$t$.

$\beta_m^{(2)}$ depends on the choice of the momentum cutoff function
$K$.  In order to get a concrete value, let us choose the step
function
\begin{equation}
K(p) \equiv \theta (1-p^2) = \left\{ \begin{array}
{r@{\quad\text{if}\quad}l} 1 & p^2 < 1\\ 0 & p^2 > 1 \end{array}
\right. \label{step}
\end{equation}
The first integral on the right-hand side of (\ref{betamgamma})
vanishes with this choice.  For the two remaining double integrals,
the integrals over $q$ can be done analytically, but the integrals
over $p$ have been done only numerically.  The final result is
\begin{equation}
\beta_m^{(2)} + 2 \gamma^{(2)} = \frac{1}{(4\pi)^4}
\end{equation}
Hence, using Eq.~(\ref{gammaresult}), we get
\begin{equation}
\beta_m^{(2)} = \frac{1}{(4\pi)^4} \frac{5}{6}
\end{equation}
This is the same result as in the minimal subtraction (MS) scheme in
dimensional regularization.  It is interesting to see whether the
equivalence of the choice (\ref{step}) to the MS scheme extends beyond
this order.

\subsection{Order $\lambda^3$}

We start from
\begin{equation}
\V_8^{(3)} (p_1,\cdots,p_8; m^2) = \frac{1 -
K(p_1+p_2+p_3)}{(p_1+p_2+p_3)^2 + m^2} \cdot \frac{1 - K(p_4 + p_5 +
p_6)}{(p_4+p_5+p_6)^2+m^2} + \text{279 permutations}
\end{equation}
By a straightforward enumeration of diagrams, we obtain
\begin{eqnarray}
&&A_8^{(3)} (q,-q,q',-q',0,0,0,0) = 18 \left\lbrace \left(
\frac{1-K(q)}{q^2} \right)^2 + \left( \frac{1-K(q')}{q'^2} \right)^2
\right\rbrace \nonumber\\
&& + 24 \left( \frac{1 - K(q)}{q^2} + \frac{1 - K(q')}{q'^2} \right)
\left( \frac{1 - K(q+q')}{(q+q')^2} + \frac{1 - K(q-q')}{(q-q')^2}
\right) \nonumber\\
&& + 24 \frac{1-K(q)}{q^2} \frac{1-K(q')}{q'^2} + 12 \left(
\frac{1-K(q+q')}{(q+q')^2} \right)^2 + 12 \left(
\frac{1-K(q-q')}{(q-q')^2} \right)^2
\end{eqnarray}
Substituting this into the integral equation for
$A_6^{(3)}(p,-p,0,0,0,0)$, we eventually obtain
\begin{eqnarray}
&&A_6^{(3)} (p,-p,0,0,0,0) = 
12 \Bigg[ \, \int_q \frac{1-K(q)}{q^2} \left(
\frac{1-K(p+q)}{(p+q)^2} \right)^2 \nonumber\\ &&\quad + \frac{1 -
K(p)}{p^2} \int_q \frac{1 - K(q)}{q^2} \left( \frac{1 -
K(p+q)}{(p+q)^2} - \frac{1 - K(q')}{q'^2} \right) \, \Bigg]
\end{eqnarray}
In getting this result, the integral over the logarithmic scale
parameter $t$ has been done by rewriting the integrand as much as
possible as a total derivative with respect to $t$.

Therefore, we obtain
\begin{eqnarray}
&&\beta^{(3)} - 4 \gamma^{(2)} = - \frac{1}{2} \int_p \frac{\Delta
(p)}{p^2} \, A_6^{(3)} (p,-p, 0,0,0,0)\nonumber\\ &&= - 6 \Bigg[ \,
\int_p \frac{\Delta (p)}{p^2} \int_q \frac{1 - K(q)}{q^2} \left(
\frac{1 - K(p+q)}{(p+q)^2} \right)^2 \\ && \, + \int_p \frac{\Delta
(p)(1 - K(p))}{p^4} \int_q \frac{1 - K(q)}{q^2} \left( \frac{1 -
K(p+q)}{(p+q)^2} - \frac{1 - K(q)}{q^2} \right) \, \Bigg]
\end{eqnarray}
This should not depend on the choice of $K$, and we are free to choose
$K$ as the step function (\ref{step}).  As before, the integrals over
$q$ can be done analytically, but the integrals over $p$ have been
done only numerically.  Our final result is
\begin{equation}
\beta^{(3)} - 4 \gamma^{(2)} = - \frac{6}{(4\pi)^4}
\end{equation}
Hence, using Eq.~(\ref{gammaresult}), we obtain
\begin{equation}
\beta^{(3)} = - \frac{17}{3} \frac{1}{(4\pi)^4}
\end{equation}
This agrees with the standard universal result.

\section{\label{appendix:improvement}Alternative definition of
the beta function and anomalous dimensions}

In Ref.~\onlinecite{hl88} Hughes and Liu define a beta function and
anomalous dimensions for the $\phi^4$ theory.  Their definition is
unsatisfactory in two respects:
\begin{enumerate}
\item It is based upon a careless treatment of counterterms. (See
the first paragraph in sect.~\ref{sec:counter}.)
\item The renormalization scheme adopted is not mass independent.
\end{enumerate}
In this appendix, we will improve their definition to come up with a
mass independent scheme with an alternative beta function and
anomalous dimensions.  We will treat counterterms carefully using the
results of sects.~\ref{sec:counter}, \ref{sec:linear}.

As has been explained in Ref.~\onlinecite{s02}, in the continuum limit
the solution $\{\V_{2n}(t)\}$ of the original ERG equation is
parametrized by four parameters: a squared mass $m^2$ and three
constants of integration $B_2 (t=0), C_2 (0), A_4 (0)$.  It is
convenient to choose the minimal subtraction scheme:
\begin{equation}
B_2 (0) = C_2 (0) = 0 \label{convention}
\end{equation}
so that the vertices $\{\V_{2n} (t)\}$ are determined by $m^2$ and
$A_4 (0)$.

As explained in sect.~\ref{sec:modify}, the problem with the above
scheme is that (\ref{convention}) is not preserved along the ERG flow.
After renormalization by an infinitesimal logarithmic scale $\Delta
t$, we have found (see Eqs.~(\ref{tildeCtwo}, \ref{tildeBtwo}))
\begin{eqnarray}
B_2 (\Delta t) &=& \Delta t\, \frac{1}{2} \int_q \Delta (q) \left(
\frac{B_4 (q,-q,0,0)}{q^2} - \frac{A_4 (q,-q,0,0)}{q^4} \right)\\ C_2
(\Delta t) &=& \Delta t\, \frac{1}{2} \frac{\partial}{\partial p^2}
\int_q \frac{\Delta (q)}{q^2} \, A_4 (q,-q,p,-p) \Bigg|_{p^2=0}
\end{eqnarray}
and (\ref{convention}) is not satisfied anymore.

We will not modify the ERG flow.  Instead we proceed by generating a
new ERG trajectory $\{\V'_{2n}(t)\}$ equivalent to the original
trajectory $\{\V_{2n} (t)\}$ in such a way that the convention
(\ref{convention}) is satisfied at $t=\Delta t$.  This can be done by
using the results of sect.~\ref{sec:linear} (see the remark in the
last paragraph).  We generate $\{\V'_{2n} (t)\}$ by introducing
generalized counterterms corresponding to the choice
\begin{equation}
u(p) = (1 - K(p)) \left( \frac{\ep m^2 \e^{2t}}{p^2 + m^2 \e^{2t}}
+ \eta \right),\quad t(p) = 0
\end{equation} 
where $\ep, \eta$ are infinitesimal constants independent of $t$.
Then, we obtain
\begin{eqnarray}
&&\V'_2 (t; p) = \ep m^2 \e^{2t} + \eta (p^2 + m^2 \e^{2t})
 \nonumber\\ &&\quad + \left( 1 - \eta + 2 (1 - K(p))
 \left(\frac{\ep m^2 \e^{2t}}{p^2 + m^2 \e^{2t}} + \eta\right) \right)
 \V_2 (t; p)\nonumber\\ &&\quad - \V_2 (t;p)^2 \frac{K(p)(1 -
 K(p))}{p^2 + m^2 \e^{2t}} \left( \frac{\ep m^2 \e^{2t}}{p^2 + m^2
 \e^{2t}} + \eta \right)\nonumber\\ &&\quad - \frac{1}{2} \int_q
 \frac{K(q) (1 - K(q))}{q^2 + m^2 \e^{2t}} \left( \frac{\ep m^2
 \e^{2t}}{q^2 + m^2 \e^{2t}} + \eta \right)\, \V_4 (t; q,-q, p, -p)
\label{primeVtwo}\\
 && \V'_{2n \ge 4} (t; p_1,\cdots,p_{2n})\nonumber\\ &&\,= \V_{2n}
 (t; p_1,\cdots, p_{2n}) \left( 1 - n \eta + \sum_{i=1}^{2n} (1 -
 K(p_i)) \left( \frac{\ep m^2 \e^{2t}}{p_i^2 + m^2 \e^{2t}} + \eta
 \right) \right) \nonumber\\ &&\quad -
 \sum_{k=0}^{\left[\frac{n-1}{2}\right]} \sum_{\text{partitions}\atop
 I+J=\{2n\}} \V_{2(k+1)} (t; p_I) \frac{K(p_I) (1 - K(p_I))}{p_I^2 +
 m^2 \e^{2t}} \left( \frac{\ep m^2 \e^{2t}}{ p_I^2 + m^2 \e^{2t}} +
 \eta \right) \V_{2(n-k)}(t; p_J)\nonumber\\ &&\quad - \frac{1}{2}
 \int_q \frac{K(q)(1-K(q))}{q^2 + m^2 \e^{2t}} \left( \frac{\ep m^2
 \e^{2t}}{q^2 + m^2 \e^{2t}} + \eta \right) \V_{2(n+1)} (t; q,-q, p_1,
 \cdots, p_{2n})
\end{eqnarray}
The new vertices $\{\V'_{2n} (t)\}$ satisfy the original ERG equation,
and they are equivalent to $\{\V_{2n}(t)\}$, i.e., they give rise to
the same Green functions:
\begin{equation}
\vev{\phi (p_1) \cdots \phi (p_{2n-1}) \phi}_{m^2 \e^{2t}, \V(t)} = (1
- n \eta) \vev{\phi (p_1) \cdots \phi (p_{2n-1}) \phi}_{m^2 (1 + \ep)
\e^{2t}, \V' (t)} \label{RGGreen}
\end{equation}
Eq.~(\ref{primeVtwo}) gives
\begin{eqnarray}
B'_2 (0) &=& \ep + \eta - \frac{\ep}{2}\int_q \frac{K(q)(1-K(q))}{q^4}
 A_4 (q,-q,0,0)\nonumber\\
&& \quad - \frac{\eta}{2} \int_q K(q)(1-K(q)) \left( \frac{B_4
(q,-q,0,0)}{q^2} - \frac{A_4 (q,-q,0,0)}{q^4} \right)\\
C'_2 (0) &=& \eta - \frac{\eta}{2} \frac{\partial}{\partial p^2} \left(
\int_q \frac{K(q)(1-K(q))}{q^2} A_4 (q,-q,p,-p) \right)\Bigg|_{p^2=0}
\end{eqnarray}
Taking $\ep, \eta$ of order $\Delta t$, we obtain 
\begin{eqnarray}
B'_2 (\Delta t) &=& B_2 (\Delta t) + B'_2 (0) \\
C'_2 (\Delta t) &=& C_2 (\Delta t) + C'_2 (0)
\end{eqnarray}
to first order in $\Delta t$.  We can choose the constants $\ep, \eta$
so that $\{\V'_{2n} (\Delta t)\}$ satisfy the convention
(\ref{convention}):
\begin{equation}
B'_2 (\Delta t) = C'_2 (\Delta t) = 0
\end{equation}

We now have two ERG trajectories: $\{\V_{2n} (t)\}$ and $\{\V'_{2n}
(t)\}$.  They are physically equivalent.  The end point $t=0$ of the
first trajectory and the end point $t=\Delta t$ of the second satisfy
the convention (\ref{convention}).  The former end point is specified
by $m^2$ and $\lambda$, and the latter by $m^2 \e^{2 \Delta t + \ep}$
and $- A'_4 (\Delta t)$.  If we introduce an alternative beta function
and anomalous dimension of the squared mass by
\begin{equation}
\tilde{\beta} (\lambda) \equiv \frac{- A'_4 (\Delta t) + A_4
(0)}{\Delta t}\, , \quad
\tilde{\beta}_m (\lambda) \equiv \frac{\ep}{\Delta t}\, ,
\end{equation}
we can specify the end point $t=\Delta t$ of the trajectory
$\{\V'_{2n}(t)\}$ by
\begin{equation}
m^2 (\Delta t) \equiv m^2 \left( 1 + \Delta t (2 + \tilde{\beta}_m
(\lambda) \right),\quad
\lambda (\Delta t) \equiv \lambda + \Delta t ~\tilde{\beta} (\lambda)
\end{equation}
\begin{figure}
\includegraphics{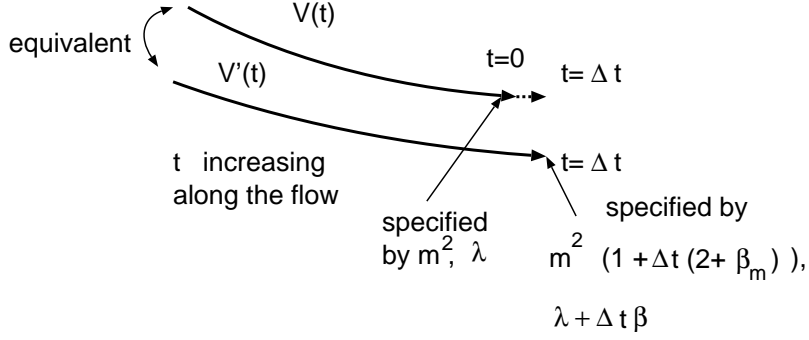}
\caption{Original ERG flows: $(m^2, \lambda)$ and $(m^2 + \Delta t
(2+\beta_m (\lambda)), \lambda + \beta (\lambda))$ specify the end
points of two different but physically equivalent ERG flows.}
\end{figure}
We can also define an alternative anomalous dimension of the scalar
field by
\begin{equation}
\tilde{\gamma} (\lambda) \equiv \frac{1}{2} \frac{\eta}{\Delta t}
\end{equation}
so that Eqs.~(\ref{RGGreen}) imply the familiar RG equation for the
Green functions:
\begin{eqnarray}
&&\vev{\phi (p_1) \cdots \phi (p_{2n-1}) \phi}_{m^2,
 \lambda}\nonumber\\ &=& (1 + (4n - y_{2n} - 2n \tilde{\gamma})\Delta
 t) \vev{\phi (p_1 \e^{\Delta t}) \cdots \phi (p_{2n-1} \e^{\Delta t})
 \phi}_{m^2 (\Delta t), \lambda (\Delta t)}
\end{eqnarray}

For completeness, let us write down the expressions of the alternative
beta function and anomalous dimensions:
\begin{eqnarray}
2 \tilde{\gamma} (\lambda) &\equiv& \frac{- \frac{1}{2}
\frac{\partial}{\partial p^2} \left(\int_q \frac{\Delta (q)}{q^2} A_4
(q,-q,p,-p) \right)_{p^2=0}}{1 - \frac{1}{2} \frac{\partial}{\partial
p^2} \left( \int_q \frac{K(q)(1-K(q))}{q^2}\, A_4 (q,-q,p,-p)
\right)_{p^2=0}}\\ \tilde{\beta}_m (\lambda) &\equiv& \frac{1}{1 -
\frac{1}{2} \int_q \frac{K(q)(1-K(q))}{q^4}\, A_4 (q,-q,0,0)}
\nonumber\\ &&\times \Bigg[ \frac{1}{2} \int_q \Delta (q) \left(
\frac{A_4 (q,-q,0,0)}{q^4} - \frac{B_4 (q,-q,0,0)}{q^2} \right)\\ && -
2 ~\tilde{\gamma} (\lambda) \left\lbrace 1 - \frac{1}{2} \int_q K(q)
(1 - K(q)) \left( \frac{B_4 (q,-q,0,0)}{q^2} - \frac{A_4
(q,-q,0,0)}{q^4}\right) \right\rbrace \Bigg] \nonumber\\ \tilde{\beta}
(\lambda) &\equiv& - \frac{1}{2} \int_q \frac{\Delta (q)}{q^2} A_6
(q,-q,0,0,0,0)\nonumber\\ && - 2 \tilde{\gamma} (\lambda) \left( 2
\lambda - \frac{1}{2} \int_q \frac{K(q)(1-K(q))}{q^2} \, A_6
(q,-q,0,0,0,0) \right)
\end{eqnarray}

The above are significantly more complicated than those
(\ref{scalegamma}, \ref{scalebetam}, \ref{scalebeta}) defined using
the modified ERG equation in sect.~\ref{sec:modify}.  But that is not
the biggest drawback of the alternative definition.  The biggest
drawback is the lack of a scaling relation for the vertices.  Let
$\{\V_{2n} (t; p_1,\cdots, p_{2n}; m^2, \lambda)\}$ denote the
vertices on the ERG trajectory whose end point at $t=0$ is specified
by $m^2, \lambda$.  Then the scaling relation would give
\begin{equation}
\V_{2n} (t; p_1, \cdots, p_{2n}; m^2, \lambda) = \V_{2n} (0; p_1,
\cdots, p_{2n}; m^2 (t), \lambda (t))
\end{equation}
But this is \textbf{NOT} the case here, simply because $\{\V_{2n} (0;
m^2 (t), \lambda(t))\}$ is not on the same ERG trajectory as
$\{\V_{2n} (0; m^2, \lambda)\}$.

\section{\label{appendix:lowest}Lowest order cutoff dependence}

In this appendix we compute $\ep (\lambda)$, $\eta (\lambda)$, $\delta
\lambda (\lambda)$, $u(p; m^2, \lambda)$, and $t(p; m^2, \lambda)$ in
sect.~\ref{sec:univ} to lowest order in $\lambda$.  To denote the
order of expansions in $\lambda$, we use the same notation as in
Appendix~\ref{appendix:beta}.

$\ep$, $\eta$, and $\delta \lambda$ are determined by
Eqs.~(\ref{etanew}, \ref{epsilonnew}, \ref{deltalambda}).  They depend
on $u$, but $u$ is first order in $\lambda$, and we do not need it for
the lowest order calculations.  We find
\begin{eqnarray}
\eta^{(2)} &=& \frac{1}{2} \frac{\partial}{\partial p^2} \int_q
\frac{\delta K(q)}{q^2} \, A_4^{(2)} (q,-q,p,-p)
\Bigg|_{p^2=0}\nonumber\\ &=& 2 \frac{\partial}{\partial p^2} \int_q
\frac{\delta K(q)}{q^2} \frac{(1 - K(q+p))}{(q+p)^2} \Bigg|_{p^2=0} =
- \int_q \frac{\delta K(q)}{q^2} K'' (q)\\ \ep^{(1)} &=& \frac{1}{2}
\int_q \frac{\delta K(q)}{q^4} ( - A_4^{(1)} (q,-q,0,0)) = -
\frac{1}{2} \int_q \frac{\delta K(q)}{q^4}\\ \delta \lambda^{(2)} &=&
\frac{1}{2} \int_q \frac{\delta K(q)}{q^2} A_6^{(2)} (q,-q,0,0,0,0) =
3 \int_q \frac{\delta K(q) (1 - K(q))}{q^4}
\end{eqnarray}
Since $\eta$ is second order in $\lambda$, we find that the anomalous
dimension $\gamma$ of the scalar field is universal (i.e., independent
of $K(p)$) up to order $\lambda^2$.

The functions $u(p; m^2, \lambda)$ and $t(p; m^2, \lambda)$ are
determined by Eqs.~(\ref{dudt}, \ref{dtdt}).  To lowest order in
$\lambda$, the equation for $u$ becomes
\begin{equation}
2 \left( p^2 \frac{\partial}{\partial p^2} + m^2
\frac{\partial}{\partial m^2} \right) u^{(1)} (p; m^2) = (\delta K(p)
\beta_m^{(1)}  + \Delta (p) \ep^{(1)} ) \frac{m^2}{p^2 + m^2}
\end{equation}
where $\beta_m^{(1)}$ is given by Eq.~(\ref{betamfirst}).  Hence, we
obtain
\begin{equation}
u^{(1)}(p; m^2) = \frac{m^2}{p^2 + m^2} \left( \beta_m^{(1)} \int_0^\infty
dt\, \delta K(p \e^{-t}) + \ep^{(1)} (1 - K(p)) \right)
\end{equation}
In fact we can add an arbitrary function $c$ of $p^2/m^2$ to the above
solution.  But $u$ must vanish for $p^2 < 1$, and this imposes $c = 0$
for $p^2 < 1$ for an arbitrary $m^2$.  Thus, we conclude $c = 0$
identically.

To lowest order in $\lambda$, the equation (\ref{dtdt}) for $t(p; m^2,
\lambda)$ gives
\begin{equation}
2 \left( p^2 \frac{\partial}{\partial p^2} + m^2
\frac{\partial}{\partial m^2} \right) t^{(2)} (p; m^2) = \beta_m^{(1)}
u^{(1)} (p;m^2) \frac{m^2}{p^2+m^2}
\end{equation}
Its unique solution is given by
\begin{equation}
t^{(2)} (p) = \beta_m^{(1)} \left( \frac{m^2}{p^2+m^2}\right)^2 \left(
\beta_m^{(1)} \int_0^\infty dt' \int_0^\infty dt \, \delta K(p
\e^{-\tpt}) + \ep^{(1)} \int_0^\infty dt (1 - K(p \e^{-t})) \right)
\end{equation}

\section{\label{appendix:SSB}Spontaneous symmetry breaking}

For $m^2 < 0$, we have spontaneous symmetry breaking, and the scalar
field acquires a non-vanishing expectation value:
\begin{equation}
\vev{\phi} = M > 0
\end{equation}

The ERG equation is still valid irrespective of the sign of $m^2$ as
long as
\begin{equation}
m^2 > - 1
\end{equation}
This is because in the ERG equation the dangerous denominator $p^2 +
m^2$ appears always as either
\begin{equation}
\frac{\Delta (p)}{p^2 + m^2} \quad \text{or}\quad \frac{1 -
K(p)}{p^2+m^2}
\end{equation}
where $\Delta (p) = 1 - K(p) = 0$ for $p^2 < 1$.

However, the perturbative calculations of the Green functions cannot
be done with the propagator
\begin{equation}
\frac{K(p)}{p^2 + m^2}
\end{equation}
because the denominator vanishes at $p^2 = - m^2 < 1$ for which $K(p)
= 1$.  Therefore, we need to rearrange the interaction terms in the
full action to get a sensible propagator.  Otherwise perturbative
calculations will not make sense.

Given a full action
\begin{eqnarray}
S[\phi] &=& \frac{1}{2} \int_p \frac{p^2 + m^2}{K(p)} \phi (p) \phi
(-p) \nonumber\\ && - \sum_{n=1}^\infty \frac{1}{(2n)!} \int_{p_1,
\cdots, p_{2n-1}} \V_{2n} (p_1,\cdots,p_{2n}) \phi (p_1) \cdots \phi
(p_{2n})
\end{eqnarray}
we introduce the replacement
\begin{equation}
\phi (p) = \varphi (p) + M (2\pi)^4 \delta^{(4)} (p)
\end{equation}
where $\varphi$ has zero expectation value.  Dropping the $\varphi$
independent constant terms, we obtain
\begin{eqnarray}
S[\phi] &=& \varphi (0) \left( m^2 M - \sum_{n=1}^\infty
\frac{M^{2n-1}}{(2n-1)!} \V_{2n} (0,\cdots,0) \right) + \frac{1}{2}
\int_p \frac{p^2 + m^2}{K(p)} \, \varphi (p) \varphi (-p) \nonumber\\
&& - \sum_{l=2}^\infty \frac{1}{l!}  \int_{p_1,\cdots,p_{l-1}} \varphi
(p_1) \cdots \varphi (p_l) \sum_{n =
\left[\frac{l+1}{2}\right]}^\infty \frac{M^{2n-l}}{(2n-1)!} \V_{2n}
(p_1,\cdots,p_l,0,\cdots,0)
\end{eqnarray}

To find the propagator, we extract the coefficient of the quadratic
term which is given by
\begin{equation}
\frac{p^2+m^2}{K(p)} - \sum_{n=1}^\infty \frac{M^{2n-2}}{(2n-2)!}
\V_{2n} (p,-p,0,\cdots,0)
\end{equation}

In calculating the Green functions in powers of the coupling constant
$\lambda$, we regard both $m^2$ and $\lambda M^2$ as order $1$.  (At
tree level, $m^2 = - \lambda M^2/6$.)  Hence, the order $1$ contribution
to the quadratic term is given by
\begin{equation}
C(p) \equiv \frac{p^2+m^2}{K(p)} - \sum_{n=2}^\infty \frac{(- \lambda
M^2)^{n-1}}{(2n-2)!}  \V_{2n}^{(n-1)} (p,-p,0,\cdots,0)
\end{equation}
where $(-\lambda)^{n-1} \V_{2n}^{(n-1)}$ is the tree-level vertex
given by
\begin{equation}
\V_{2n}^{(n-1)} (p,-p,0,\cdots,0) = \frac{(2(n-1))!}{2^{n-1}} \left(
\frac{1 - K(p)}{p^2+m^2} \right)^{n-2}
\end{equation}
\begin{figure}
\includegraphics{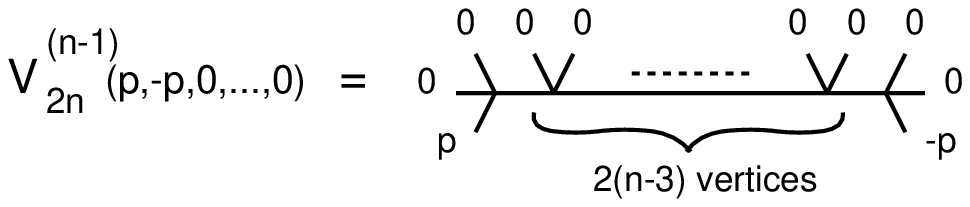}
\caption{\label{fig4}$2n$-point vertex at tree-level}
\end{figure}
(See FIG.~\ref{fig4}.)  Note $\V_2 = 0$ at tree-level.  Therefore, we
obtain
\begin{eqnarray}
C(p) &=& \frac{p^2 + m^2}{K(p)} + \frac{\lambda M^2}{2} \sum_{n=2}^\infty
\left( \frac{- \lambda M^2}{2} \frac{1 - K(p)}{p^2+m^2}
\right)^{n-2}\nonumber\\
&=& \frac{\left(p^2 + m^2 + \frac{\lambda M^2}{2}\right)
(p^2+m^2)}{K(p) \left( p^2 + m^2 + \frac{\lambda M^2}{2} (1 - K(p))
\right)}
\end{eqnarray}
Hence, the propagator is given by
\begin{eqnarray}
\frac{1}{C(p)} &=& \frac{K(p)}{p^2 + m^2 + \frac{\lambda M^2}{2}}
\cdot \frac{p^2 + m^2 + \frac{\lambda M^2}{2}(1 - K(p))}{p^2 +
m^2} \label{Cinv}\\
&=& \frac{1}{p^2 + m^2 + \frac{\lambda M^2}{2}} \quad \text{for}\quad
p^2 < 1 \nonumber
\end{eqnarray}
This is well defined for any $p^2$, and vanishes for $p^2 > 2^2$ due
to $K(p)$ in the numerator.

Therefore, the full action can be written as
\begin{equation}
S[\phi] = \frac{1}{2} \int_p C(p) \varphi (p) \varphi (-p) +
S'_{\text{int}} [\varphi]
\end{equation}
where the interaction part is given by
\begin{eqnarray}
S'_{\text{int}} [\varphi] &=& \varphi (0) \left( m^2 M -
\sum_{n=1}^\infty \frac{M^{2n-1}}{(2n-1)!}  \V_{2n} (0,\cdots,0)
\right) \nonumber\\&& - \frac{1}{2} \int_p \varphi (p) \varphi (-p) \,
\sum_{n=1}^\infty \frac{M^{2(n-1)}}{2(n-1)!} \left( \V_{2n}
(p,-p,0,\cdots,0) - \V_{2n}^{(n-1)} (p,-p,0,\cdots,0)
\right)\nonumber\\ && - \sum_{l=3}^\infty \frac{1}{l!}
\int_{p_1,\cdots,p_{l-1}} \varphi (p_1) \cdots \varphi (p_l) \sum_{n =
\left[\frac{l+1}{2}\right]}^\infty \frac{M^{2n-l}}{(2n-1)!} V_{2n}
(p_1,\cdots,p_l,0,\cdots,0)
\end{eqnarray}
This is at least of order $\lambda$, and hence perturbative
calculations can be done using the propagator (\ref{Cinv}) and the
above interaction action.

\newpage
\begin{acknowledgments}
The present work was done while the author was visiting the Department
of Physics at Penn State University.  He thanks Profs.~J.~Banavar and
M.~G\"unayden for hosting his visit.  This work was partially
supported by the Grant-In-Aid for Scientific Research from the
Ministry of Education, Culture, Sports, Science, and Technology, Japan
(\#14340077).
\end{acknowledgments}

\bibliography{beta}

\end{document}